\documentclass{aa}  

\usepackage{graphicx}

\usepackage{txfonts}

\usepackage[tight]{units}
\usepackage[acronym]{glossaries}
\usepackage{amsmath}
\usepackage{multirow}
\usepackage{enumerate}
\usepackage{soul}
\usepackage[table,dvipsnames,svgnames]{xcolor}
\usepackage{pifont}
\usepackage{lscape}
\usepackage{xspace}

\definecolor{cgood}{HTML}{c6e8cd}
\definecolor{cokish}{HTML}{fff2cc}
\definecolor{cbad}{HTML}{f4cccc}

\newacronym{SRMHD}{SRMHD}{special-relativistic magnetohydrodynamic}
\newacronym{EOS}{EOS}{equation of state}
\newacronym{GW}{GW}{gravitational wave}
\newacronym{NSE}{NSE}{nuclear statistical equilibrium}
\newacronym[shortplural={CCSNe}, longplural={core collapse supernovae}]{CCSN}{CCSN}{core collapse supernova}
\newacronym{PNS}{PNS}{proto-neutron star}
\newacronym{SASI}{SASI}{stationary accretion shock instability}
\newacronym{1D}{1D}{1-dimensional}
\newacronym{2D}{2D}{2-dimensional}
\newacronym{3D}{3D}{3-dimensional}
\newacronym{BNS}{BNS}{binary neutron star}
\newacronym{ET}{ET}{Einstein Telescope}
\newacronym{CE}{CE}{Cosmic Explorer}
\newacronym{EEMD}{EEMD}{ensemble empirical mode decomposition}
\newacronym{EMD}{EMD}{empirical mode decomposition}
\newacronym{IF}{IF}{instantaneous frequency}
\newacronym{IMF}{IMF}{intrinsic mode function}
\newacronym{sIMF}{sIMF}{significant IMF}
\glsdisablehyper
\usepackage{hyperref}

\begin{document} 

\title{Convection signatures in early-time gravitational waves from core-collapse supernovae}

\subtitle{}

\author{M.~Cusinato
        \inst{1}\thanks{marco.cusinato@uv.es}
        \and
        M.~Obergaulinger
        \inst{1}\thanks{martin.obergaulinger@uv.es}
        \and
        M.Á.~Aloy
        \inst{1}\inst{2}\thanks{miguel.a.aloy@uv.es}
        }
\institute{
 Departament d'Astronomia i Astrof\'{\i}sica, Universitat de Val\`encia, Dr. Moliner, 50, 46100 Burjassot, Spain
 \and
 Observatori Astronòmic, Universitat de València, 46980 Paterna, Spain
}

\date{}

\abstract
   {
   \Acrlongpl{GW} emitted from \acrlong{CCSN} explosions are critical observables for extracting information about the dynamics and properties of both the progenitor and the post-bounce~evolution of the system. They are prime targets for current interferometric searches and represent a key milestone for the capabilities of next-generation interferometers.
   }
   {
   This study aims  to characterise how the gravitational waveform associated with prompt stellar convection depends on the rotational rate and magnetic field topology of the progenitor star.
   }
   {
   We carried out a series of axisymmetric simulations of a $\unit[16.5]{M_\odot}$ red supergiant with five configurations of initial magnetic fields and varying degrees of initial rotation. We analysed the contribution of early-time convection and the \acrlong{PNS} core to the waveform using \acrlong{EEMD}, alongside spectral and Fourier analyses, to facilitate the comparison and interpretation of the results.
   }
    {
    Our simulations show that the first six \acrlongpl{IMF} dominate the early post-bounce \acrlong{GW} signal, with variations due to rotation and magnetic fields influencing the signal strength. Strong magnetic fields decelerate core rotation, affecting mode excitation. Regardless of the initial rotation, convection consistently drives a low-frequency mode that lasts throughout the evolution.
    }
{We conclude that prompt convection can produce \acrlong{GW} amplitudes comparable to or exceeding those of core bounce, with a persistent low-frequency component detectable in next-generation detectors.
}

\keywords{Supernovae --- Gravitational waves --- MHD --- Instabilities --- Convection}

\maketitle
\nolinenumbers
\section{Introduction} \label{sec:intro}

\Glspl{CCSN} mark the terminal evolutionary stage of massive stars ($M_\textnormal{ZAMS}\gtrsim \unit[8]{M_\odot}$).
Throughout their lives, these stars undergo nuclear fusion, creating heavier elements up to the iron group. This process results in an onion-like structure, with an iron core at the centre, primarily supported by the pressure of relativistic degenerate electrons. When the core accumulates enough mass to exceed the Chandrasekhar limit, it begins to collapse \citep[see e.g.~][]{Colgate66, Bethe90, Mezzacappa2005,Kotake2006,Janka2007, Janka12, Burrows13, Melson2015,Lentz2015,Janka2016,Mueller2016,Mueller_2020LRCA....6....3,Burrows2021,Yamada2024}. The density then rapidly increases, surpassing nuclear levels, setting the stage for the formation of a \gls{PNS}. At this point, the collapse abruptly stops, causing the external layers, which are falling supersonically towards the stellar centre, to bounce and form a shock wave that propagates outwards into the infalling matter. Photodissociation of iron-group nuclei causes the shock to lose part of its energy, causing it to stall at $\sim\unit[150]{km}$. Various mechanisms have been proposed to explain the shock revival. The neutrino mechanism \citep{Wilson1982, Bethe85}, whereby $5-10\%$ of the total outgoing neutrino luminosity from the \gls{PNS} is deposited in the region behind the shock, is sufficient to revive the shock and drive the explosion in most cases. On the other hand, a small fraction of \glspl{CCSN} can explode magneto-rotationally. In this latter case, fast rotation and strong magnetic fields enhance the energy deposition behind the shocked region, contributing to the creation of bipolar jets that drive an extremely energetic explosion \citep{Bisnovatyi-Kogan_1976Ap&SS..41..287, Mueller_1979A&A....80..147, Symbalisty_1984ApJ...285..729, Akiyama_2003ApJ...584..954, Kotake_2004ApJ...608..391, Moiseenko_2006MNRAS.370..501, Obergaulinger_2006A&A...450.1107, Burrows_2007ApJ...664..416, Dessart_2007ApJ...669..585, Winteler_2012ApJ...750L..22, Sawai_2016ApJ...817..153, Mosta14, Obergaulinger17, Bugli_2020MNRAS.492...58}.

Regardless of being the main driver of the explosion, neutrino emission and absorption in the post-shock region drive convection \citep{Colgate_1993PhR...227..157, Herant_1994ApJ...435..339, Fryer_2007ApJ...659.1438}, which is suspected to create infalling funnels over the \gls{PNS} and excite its oscillation modes \citep{Vartanyan23a}. Convection within the \gls{PNS} also seeds these oscillations \citep{Andresen17, Mezzacappa2020, Mezzacappa2023, Murphy2025}. Even though these explosion mechanisms have been extensively corroborated via theory and both \gls{2D} and \gls{3D} simulations \citep{Mosta14, Obergaulinger14a, Matsumoto22, Vartanyan23b, Wang24, Matsumoto24}, the direct detection of the multi-messenger \gls{CCSN} signature in electromagnetic, neutrino, and gravitational radiation would mark a significant advancement in our understanding of these events \citep{Kotake12}.

The current generation detectors such as Advanced LIGO \citep{Aasi15}, Advanced Virgo \citep{Acernese15}, and KAGRA \citep{Akutsu19} may detect the \glspl{GW} emitted by a galactic explosion, and even more so with the next-generation ones such as the \gls{ET} \citep{Maggiore20} and \gls{CE} \citep{Reitze19}. 
\Gls{GW} detection offer a robust means of constraining the nuclear \gls{EOS} \citep{Malik18, Raaijmakers20}, as has been demonstrated by the detection of the first \gls{BNS} merger event, GW170817 \citep{Abbott17}. Beyond this \gls{GW} also carry valuable information about the internal matter evolution during \gls{CCSN} \citep{Kuroda16,Richers17,EggenbergerAndersen2021,Murphy2024}, making them one of the most valuable tools for studying the dynamics of the inner properties of these events.

Because of the stochasticity of their generation, \gls{GW} signals from \glspl{CCSN} pose significant challenges for analysis. Many processes contribute to their emission, from the oscillations caused by prompt convection in the early post-bounce~phase, with frequencies of a few hundred Hertz \citep{Marek09, Murphy09, Yakunin10, Muller13, Yakunin15, Pan18}, to the $f$- and $g$-mode ramp-up signals whose frequencies rise to $\sim\unit[1]{kHz}$, driven by surface gravity modes of the \gls{PNS} \citep{TorresForne18, TorresForne19, Bugli23}, and finally to $\sim\unit[100]{Hz}$ features associated with the \gls{SASI} \citep{Blondin2003, Blondin2007, CerdaDuran13, Kuroda16, Andresen17, Mezzacappa2020,Mezzacappa2023}. 
These processes are physically separated and provide complementary information: prompt convection is transient and stochastic, whereas later \gls{SASI} or \gls{PNS}-mode signals are coherent, quasi-periodic phenomena reflecting the longer-term hydrodynamic and structural evolution of the supernova core.

In this context, efforts to characterise \glspl{GW} from \glspl{CCSN} date back to the early 1980s \citep{Muller82, Zwerger97}, and have revealed a wide diversity of waveforms \citep[see e.g.~][]{Dimmelmeier02, Nakamura16, Mezzacappa2023, Powell23}. Finally, \cite{Abdikamalov14} and \cite{Richers17} demonstrated that the \gls{GW} bounce signal is primarily determined by the ratio of the rotational kinetic energy to gravitational energy ($T/|W|$) of the core at bounce.

In this study, we aim to characterise the \gls{GW} signal associated with prompt convection in the early post-bounce phase as a function of the magneto-rotational properties of the collapsing core, as well as the influence of longer-term convective instabilities on \glspl{GW} emission.
Specifically, we explore how variations in rotational velocity and magnetic field strength affect the frequency, amplitude, and morphology of the resulting \gls{GW} signal. Through systematic variation of these parameters in our simulations, we aim to identify correlations between  physical core conditions and the emitted \gls{GW} features.

This paper is organised as follows. Section \ref{sec:mat_met} outlines our numerical set-up and presents the progenitor model and nuclear \gls{EOS} used to perform the \gls{CCSN} simulations. Section \ref{sec:sim_ove_analysis} describes the parameters used for the simulations and summarises the analysis methods employed in this work. In Section \ref{sec:results}, we describe and present the main findings of our study. Section \ref{sec:discussion} discusses how the detectability varies with different rotation rates and magnetic field configurations. Finally, in Section \ref{sec:conclusion} we draw our conclusions.

\section{Model and numerical set-up}
\label{sec:mat_met}

The axisymmetric simulations presented in this work were performed with the \texttt{Aenus-ALCAR} code~\citep{Obergaulinger08, Just15, Just18}, which solves the equations of \gls{SRMHD} coupled with a multi-group neutrino transport two-moments scheme. To treat densities exceeding $\unit[8\cdot 10^7]{g/cm^{3}}$, we employed the SFHo \gls{EOS}~\citep{Steiner13} for dense nuclear matter. This \gls{EOS} is broadly consistent with current astrophysical constraints, including those derived from NICER observations~\citep{Miller19, Miller21, Riley19, Riley21} and \gls{GW} events~\citep{Abbott17, LVK19}. Moreover, it accounts for neutrons, protons, electrons, positrons, and photons, as well as light nuclei (such as deuterium, tritium, and helium) and heavy nuclei in \gls{NSE}.

The effects of neutrinos during the simulation are treated in the two-moment framework closed by the maximum-entropy Eddington factor~\citep{Cernohorsky94}, with the inclusion of gravity in the neutrino transport equation following the $\mathcal{O}(v/c+)$ formulation presented in~\cite{Endeve12}. Neutrino-matter interactions consist of nucleonic absorption, emission, and scattering with corrections due to weak magnetism and recoil; nuclear absorption, emission, and scattering; inelastic scattering off electrons; pair processes (electron–positron annihilation), and nucleonic bremsstrahlung.

All the \gls{CCSN} simulations presented in this work follow the evolution of the red supergiant stellar model \verb|s16.5|. This model results from the spherically symmetric evolution of a progenitor star with $M_\textnormal{ZAMS}=\unit[16.5]{M_\odot}$~\citep{Sukhbold14}, solar metallicity, no rotation, and no magnetic fields. At the pre-collapse~phase, the star retains only $\unit[14]{M_\odot}$, and the compactness of the core at $\unit[2.5]{M_\odot}$\footnote{Defined following \cite{Oconnor2011}: $\xi_{M}=\frac{M/M_\odot}{R(M_\textnormal{bary}=M)/\unit[1000]{km}}$} is 0.16.

Since the considered model is non-rotating, we superimposed an ad hoc rotation profile on the pre-collapse~star~\citep{Eriguchi_1985A&A...146..260}:
\begin{equation}
    \label{eq:rotation_prof}
    \Omega(r)=\Omega_0\left(\frac{r_0}{r + r_0}\right)^2.
\end{equation}
The radius $r_0$ marks the location at which rigid ($r < r_0$) transitions into differential rotation ($r > r_0$).
We set $r_0 = \unit[1000]{km}$ in all models. $\Omega_0$ is the maximum initial rotation rate, and $r$ is the spherical radius.

In a similar fashion, we added magnetic fields by initially introducing toroidal, $B_\textnormal{tor}$, and poloidal, $B_\textnormal{pol}$, components at the pre-collapse phase of the supernova. We then define a field geometry with the vector potential $\mathbf{A}$ characterised by a typical distance $R_0=\unit[2\cdot10^8]{cm}$, \citep{Suwa07}:
\begin{equation}
    \label{eq:vector_potential}
        \left(A^r, A^\theta, A^\phi\right) = \frac{R_0^3r}{2\left(r^3+R_0^3\right)}\big(B_\textnormal{tor}\cos\theta, 0, B_\textnormal{pol}\sin\theta\big).
\end{equation}

All the models were simulated using spherical grids consisting of $n_\theta=128$ zones in the $\theta$ direction, covering the whole polar domain $\theta\in[0,\,\pi]\,\textnormal{rad}$ with a resolution of $1.4^\circ$. In the radial direction, we employ a logarithmically stretched grid with $n_\textnormal{r}=480$ zones that extend from the centre of the domain to $\unit[10^{10}]{cm}$. This logarithmic stretching ensures an equal aspect ratio (i.e.~ $\Delta r = r\Delta\theta$) of the cells down to a uniform resolution of $\unit[4\cdot10^4]{cm}$ in the centre of the computational domain.
In the energy domain, we used a logarithmically spaced grid with $n_\epsilon=12$ energy bins ranging from $\epsilon_\textnormal{min}=\unit[0]{MeV}$ to $\epsilon_\textnormal{max}=\unit[440]{MeV}$.

Two models  were also run at higher resolution ($n_r=480$, $n_\theta=256$). Their results remain consistent with the standard-resolution runs, differing by less than $\sim26\%$ in the quantitative values of Section~\ref{sec:bounce_convection}.

\section{Simulations overview and analysis methods}
\label{sec:sim_ove_analysis}

\subsection{Simulations overview}
\label{sec:sim_overview}

We performed 29 \gls{CCSN} simulations starting from the red supergiant progenitor, \texttt{s16.5} (see Section \ref{sec:mat_met}).
In our models, we process the \gls{GW} emission at run-time, producing outputs of the \gls{GW}-strain and associated variables every $\unit[0.01]{ms}$. 
We add rotation and magnetic field profiles using Equations (\ref{eq:rotation_prof}) and (\ref{eq:vector_potential}), respectively. The range of central, pre-collapse~rotation rate, $\Omega_0$, is $\left[0,\,2.4\right]\unit[]{\textnormal{rad}\cdot s^{-1}}$, while the toroidal and poloidal components of the magnetic field span $\left[0,\,5\right]\cdot\unit[10^{11}]{G}$ and $\left[0,\,1\right]\cdot\unit[10^{12}]{G}$. The values of the rotational rate and magnetic field  strength are within the range usually encountered for models of massive stars in rapid rotation \citep[e.g.][]{Woosley_2006ApJ...637..914, Griffiths_2022A&A...665A.147}. 
Models in  Table \ref{tab:table_1} are named following the format \texttt{sA.B-C}, where \texttt{A.B} denotes the value of $\Omega_0$ in $\unit[]{\textnormal{rad}\cdot s^{-1}}$, and \texttt{C} can be equal to 0, 1, 2, 3, or 4, reflecting increasing initial magnetic field strength (\texttt{C}=\texttt{0} denotes models without magnetic field).

We output data every $\unit[1]{ms}$, recording key physical quantities to analyse the link between the dynamics and the \gls{GW} signal. For a refined analysis, models \texttt{s0.0-1}, \texttt{s1.0-1}, and \texttt{s2.0-1} were rerun from $\unit[10]{ms}$ before bounce to at least $\unit[100]{ms}$ after, with an output cadence of $\unit[0.1]{ms}$.

\begin{table}[t]
    \caption{\label{tab:table_1} List of models.}
    \centering
    \begin{tabular}{ccccc}
             \multirow{2}{*}{Model name}	& $\Omega_0$ 	&	$B_\textnormal{pol}$ 	&	$B_\textnormal{tor}$ 	&	$t_\textnormal{end}$	\\
             & $\left[\textnormal{rad}\cdot s^{-1}\right]$ & [$\unit[10^8]{G}$] & [$\unit[10^8]{G}$] & [s] \\\hline
        \texttt{s0.0-1} & 0.0 & $5$ & $10^1$ & 0.564 \\ 
        \texttt{s0.2-1} & 0.2  & $5$ & $10^1$ & 0.604 \\ 
        \texttt{s0.4-1} & 0.4  & $5$ & $10^1$ & 0.481 \\ 
        \texttt{s0.6-0} & 0.6  & $0$ & $0$ & 0.51 \\ 
        \texttt{s0.6-1} & 0.6  & $5$ & $10^1$ & 0.485 \\ 
        \texttt{s0.6-2} & 0.6  & $5\cdot 10^1$ & $10^2$ & 0.418 \\ 
        \texttt{s0.6-3} & 0.6  & $5\cdot 10^2$ & $10^3$ & 0.381 \\ 
        \texttt{s0.6-4} & 0.6  & $5\cdot 10^3$ & $10^4$ & 0.736 \\ 
        \texttt{s0.8-1} & 0.8  & $5$ & $10^1$ & 0.45 \\ 
        \texttt{s0.9-1} & 0.9  & $5$ & $10^1$ & 0.603 \\ 
        \texttt{s0.95-1} & 0.95  & $5$ & $10^1$ & 0.45 \\ 
        \texttt{s1.0-0} & 1.0  & $0$ & $0$ & 0.27 \\ 
        \texttt{s1.0-1} & 1.0  & $5$ & $10^1$ & 0.48 \\ 
        \texttt{s1.0-2} & 1.0  & $5\cdot 10^1$ & $10^2$ & 0.439 \\ 
        \texttt{s1.0-3} & 1.0  & $5\cdot 10^2$ & $10^3$ & 0.29 \\ 
        \texttt{s1.0-4} & 1.0  & $5\cdot 10^3$ & $10^4$ & 0.478 \\ 
        \texttt{s1.05-1} & 1.05  & $5$ & $10^1$ & 0.484 \\ 
        \texttt{s1.1-1} & 1.1  & $5$ & $10^1$ & 0.352 \\ 
        \texttt{s1.2-1} & 1.2  & $5$ & $10^1$ & 0.565 \\ 
        \texttt{s1.4-1} & 1.4  & $5$ & $10^1$ & 0.495 \\ 
        \texttt{s1.6-1} & 1.6  & $5$ & $10^1$ & 0.574 \\ 
        \texttt{s1.8-0} & 1.8  & $0$ & $0$ & 0.467 \\ 
        \texttt{s1.8-1} & 1.8  & $5$ & $10^1$ & 0.443 \\ 
        \texttt{s1.8-2} & 1.8  & $5\cdot 10^1$ & $10^2$ & 0.547 \\ 
        \texttt{s1.8-3} & 1.8  & $5\cdot 10^2$ & $10^3$ & 0.347 \\ 
        \texttt{s1.8-4} & 1.8  & $5\cdot 10^3$ & $10^4$ & 0.514 \\ 
        \texttt{s2.0-1} & 2.0  & $5$ & $10^1$ & 0.347 \\ 
        \texttt{s2.2-1} & 2.2  & $5$ & $10^1$ & 0.286 \\ 
        \texttt{s2.4-1} & 2.4  & $5$ & $10^1$ & 0.595 \\\hline
    \end{tabular}
    \tablefoot{Columns from left to right: name of the simulation, pre-collapse maximum rotational frequency (Equation~(\ref{eq:rotation_prof})); $B_{\rm pol}$ and $B_{\rm tor}$ (Equation~(\ref{eq:vector_potential})), and simulated time elapsed from core bounce.}
\end{table}

\subsection{Analysis method}
\label{sec:analysis_method}

In this section we describe the methodology to identify contributions to the GW signal from different locations in our models (Section \ref{sec:GW_zones}), the mathematical tool employed for the analysis of the \gls{GW} signal, based upon \gls{EEMD} (Section \ref{sec:EEMD}), and the criterion employed to identify convective zones (Section \ref{sec:conv}).

\subsubsection{Identification of the GW emission regions}
\label{sec:GW_zones}

The use of the approximate quadrupole formula (see App.~\ref{sec:GWextraction}) allows us to identify contributions to the full \gls{GW} signal by restricting the integral to subdomains inside the core. In the literature, similar decompositions have been employed, but these have mainly focused on subdividing the \gls{PNS} interior itself \citep{Andresen2017,Mezzacappa2023,Murphy2025}. In contrast, here we compute the partial signals coming from three regions:
(i) the \gls{PNS} core (inside the \gls{PNS}), (ii) the convective `sonic envelope' (outside the \gls{PNS}), and (iii) the outer region.

The actual definitions of these three regions follow.
We define the \gls{PNS} core as the  region within the \gls{PNS} where the entropy per baryon is smaller than $\unit[4]{k_B/bry}$. This value is commonly used in the literature to diagnose the explosion properties of pre-collapse stellar models \citep[e.g.~][]{Ertl16}.
The sonic envelope is the region in sonic contact with the centre (where $|v_r|<c_s$) excluding the \gls{PNS} core. Moreover, to account for convective effects associated with rotationally driven matter transport, we extend the outer boundary of the sonic envelope by $\unit[20]{km}$.
Finally, the outer region is defined as the domain beyond the convective shell.
The definitions of the above division are motivated by the characteristic frequencies of the resulting \glspl{GW} emitted from each region as well as by the fundamentally different matter regimes in them (see also Section \ref{sec:qualitative}).
Hereafter we will refer to the \glspl{GW} emitted from the \gls{PNS} core as `core strain', from the sonic envelope as `sonic envelope strain', and from the stellar region outside the sonic envelope as `outer strain'.

\subsubsection{Ensemble empirical mode decomposition method}
\label{sec:EEMD}

The \gls{EMD} is a method of decomposing a complex signal into a finite set of \glspl{IMF}, which represent simple oscillatory modes, as proposed by \cite{Huang98, Huang99}.
Considering a signal, $s(t)$, the first step of the algorithm involves identifying all the local maxima and minima within the signal. Using these points, an upper (lower) envelope is constructed by interpolating between the maxima (minima).

Once the envelopes have been calculated, their average is computed and then subtracted from the original signal. The result serves as a candidate for the \gls{IMF}. For a component to qualify as an \gls{IMF}, it must satisfy two conditions: the number of extrema and zero crossings must either be equal or differ by at most one, and the mean value of the envelope defined by the local maxima and minima must be zero at any point \citep{Flandrin05}.

The process, called `sifting', is iterated by treating the \gls{IMF} candidate as a new signal.
The sifting continues until the \gls{IMF} candidate changes negligibly between successive iterations or other stopping criteria are met, such as a predefined number of sifting iterations or a sufficiently low energy level in the residual signal.

After an \gls{IMF} is extracted, it is subtracted from the original signal to obtain the residual. The residual then becomes the new signal, and the entire process is repeated to extract the subsequent \glspl{IMF}. The algorithm stops when the residual is reduced to a monotonic function or one with no more extrema.

The \gls{EMD} process results in the decomposition of the original signal as
\begin{equation}
    s(t) = \sum_{i=0}^k c_i(t) + r(t),
\end{equation}
where $c_i(t)$ are the \glspl{IMF} ($k+1$ in the previous equation) and $r(t)$ is the residual.

One of the major shortcomings of this method is mode mixing. To address this, \cite{Wu09} proposed the \gls{EEMD}.
The algorithm is modified by adding white Gaussian noise to the original signal before calculating the \glspl{IMF} and repeating this process $n$ times. Finally, each ensemble is averaged to get the final \glspl{IMF}.
Thus two additional parameters are introduced: the ratio of the standard deviation of the Gaussian white noise to that of the original signal, $\sigma_{\textnormal{eemd}}$, and the number of ensemble trials, $n$.
By doing so, mode mixing is eliminated (or significantly reduced for a finite $n$), while the added white noise is averaged out through the ensemble mean.

The procedure we employed to decompose the \gls{GW} signal into \glspl{IMF} is as follows:
\begin{enumerate}[i.]
    \item Filter the signal to remove all frequencies above \unit[3000]{Hz};
    \item Perform standard \gls{EMD} to obtain the \glspl{IMF} and a residual;
    \item Remove the residual from the filtered signal;
    \item Perform \gls{EEMD}, limiting the number of \glspl{IMF} to 10, with $\sigma_0 = 1$ and $n = 2 \times 10^6$.
\end{enumerate}

To calculate the \gls{EMD} and \gls{EEMD}, we used the \texttt{PyEMD} package \citep{pyemd} and the \texttt{akima} interpolation method \citep{Akima1970} to find the extrema.

Additionally, it is also possible to extract the \gls{IF} of a single \gls{IMF}, a sum of them, or the entire signal. The procedure, described in \cite{Huang98b}, consists of applying the Hilbert transform to the signal (or \gls{IMF}). This allows for the construction of a composite signal whose real and imaginary components are the original signal and the Hilbert transform, respectively. Finally, by representing this complex signal in polar form, it is possible to determine its instantaneous phase, $\Phi(t)$, and to define the \gls{IF} as

\begin{equation}
\label{eq:IF}
    \omega(t) = \frac{\textnormal{d}\,\Phi(t)}{\textnormal{d}t}.
\end{equation}

\subsubsection{Identification of convective regions} 
\label{sec:conv}

To locate convective zones in our models, we used the sign of the squared Brunt-Väisälä frequency, $N^2$ \citep{Obergaulinger_2009A&A...498..241, Gossan20},
\begin{equation}
    \label{eq:BV_freq}
    N^2=\frac{\partial \Phi}{\partial r} \frac{1}{\rho}\left(\frac{1}{c_s^2}\frac{\partial P}{\partial r} - \frac{\partial \rho}{\partial r} \right),
\end{equation}
where $c_s^2$ is the speed of sound, $\Phi$ the gravitational potential, and $P$ the gas pressure. We evaluated $N^2$ for each angle, $\theta$, separately and identified convective zones by $N^2<0$.

Additionally, we estimated the modulus of the convective velocity as
\begin{equation}
    \label{eq:conv_vel}
    v_{\textnormal{conv}} = \sqrt{(v_r - \langle v_r\rangle_\Omega)^2 + (v_\theta- \langle v_{\theta}\rangle_\Omega)^2 }
,\end{equation}
where $\langle v_i\rangle_\Omega$ is the angular average of the velocity component $i= r, \theta$, defined as
\begin{equation}
    \label{eq:ang_average}
    \langle v_{i}\rangle_\Omega = \frac{\int\textnormal{d}\Omega \, v_i \rho}{\int\textnormal{d}\Omega \, \rho}.
\end{equation}
Finally, from Equation (\ref{eq:conv_vel}) it is possible to derive the associated convection frequency as 
\begin{equation}
    \label{eq:conv_f}
    f_{\textnormal{conv}} = \frac{v_{\textnormal{conv}}}{r}.
\end{equation}

\section{Results}
\label{sec:results}

To quantify the degree of rotation, we employed the sum of the rotational kinetic energies of the \gls{PNS} core and sonic envelope divided by their gravitational energy, i.e. to the ratio $T/|W|$ at core bounce.
Specifically, in the following sections we identify models with $T/|W|\leq0.004$ as slowly rotating, with $0.004\leq T/|W|\leq0.010$ as intermediately rotating\footnote{Model \texttt{IR} of \cite{Cusinato25} belongs to this class.}, with $0.010\leq T/|W|\leq0.022$ as fast-rotating, and finally $T/|W|\geq 0.022$ as very fast-rotating. The results of our simulations are summarised in Table \ref{tab:results}.

\subsection{Qualitative overview}
\label{sec:qualitative}

\begin{figure*}[t]
\centering
    \includegraphics[width=17cm]{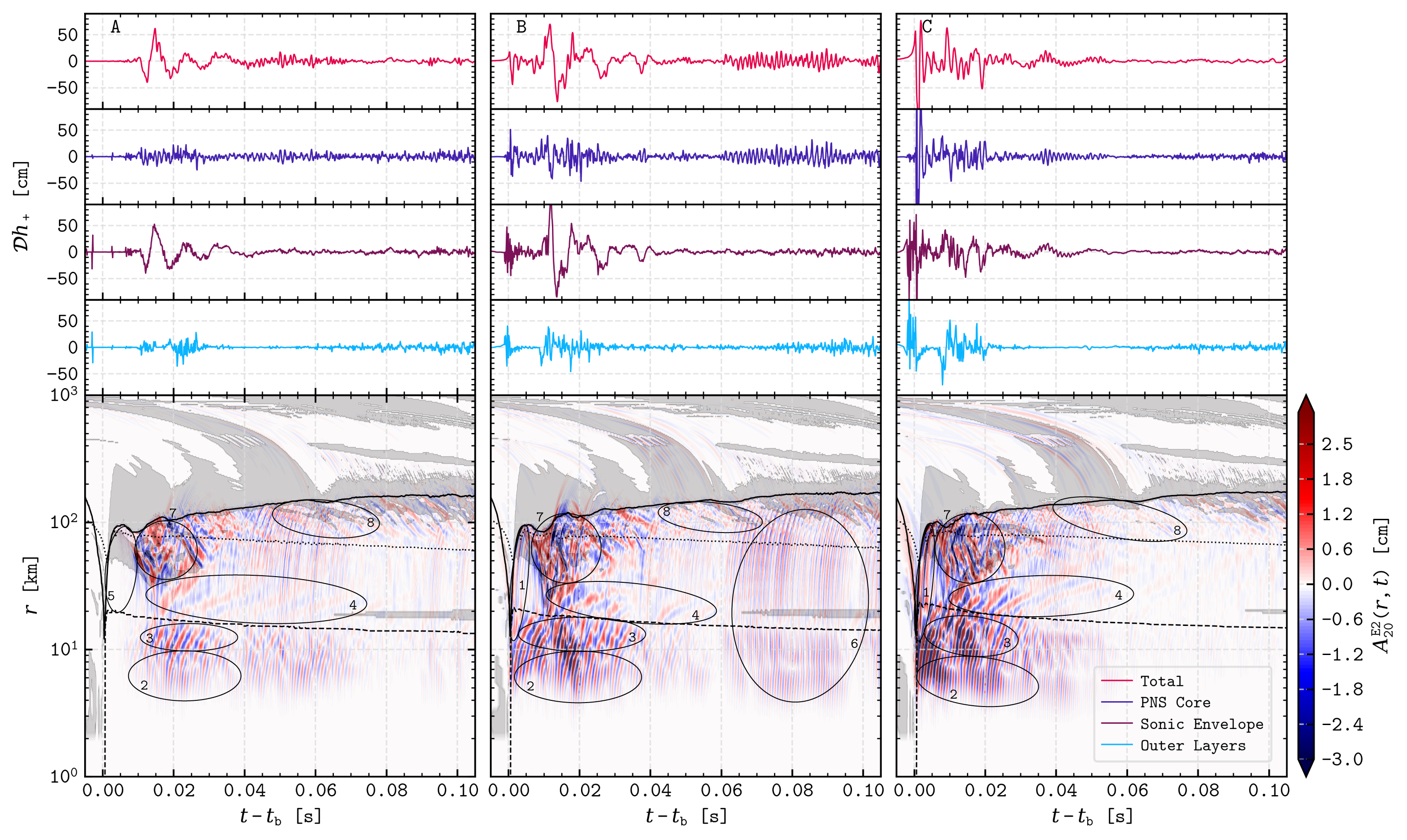}
    \caption{Evolution of \gls{GW} amplitude for models \texttt{s0.0-1} (A), \texttt{s1.0-1} (B), and \texttt{s2.0-1} (C). Panels from top to bottom show the time evolution of the \gls{GW} amplitude for the whole simulation domain, \gls{PNS} core, sonic envelope, and outer layers. Finally, the bottom row shows the space-time evolution of $\mathcal{D}h(t, r)$. Dashed, solid, and dotted lines represent the \gls{PNS} core, sonic envelope, and \gls{PNS} average radius, respectively. Grey shades mark  regions  where $N^2<0$, i.e.~ approximately, regions where convection takes place. Numbered ellipses identify \gls{GW} modes in each region (see text).
    } 
    \label{fig:decomposition}
\end{figure*}

We first present a qualitative overview of the \gls{GW} emission during the first $\unit[100]{ms}$ of post-bounce evolution, focusing on core bounce, \gls{PNS} core vibrations, and convection. This analysis covers three representative pre-collapse rotation profiles: non-rotating (\texttt{s0.0-1}), intermediate (\texttt{s1.0-1}), and fast-rotating (\texttt{s2.0-1}). Unless otherwise noted, discussion of mode frequencies and spatial distributions refers to the last row of Figure~\ref{fig:decomposition}, while the total, \gls{PNS} core, sonic envelope, and outer strains are shown in the first, second, third, and fourth rows, respectively. In brackets we indicate the label of features in the last row of Figure~\ref{fig:decomposition} that can be associated with specific modes or other phenomena in the \gls{GW} strain.

\subsubsection{\label{sec:qualitative_bounce}Core bounce}

    The \gls{PNS} forms and abruptly halts the infalling matter, inducing vibrations in the core. At this stage, only the rotating models are sufficiently aspherical to produce significant \gls{GW} emission. The emission cannot be neatly attributed to a single post-shock region. Instead, it originates from matter between $r \gtrsim \unit[12]{km}$ and the rapidly expanding, centrifugally deformed \gls{PNS} surface, which, during this brief phase, coincides with the shock wave (1).
  This spatially extended emission produces the well-known pattern of an initial rise, a sharp minimum, and a subsequent maximum, lasting in total less than $\approx \unit[5]{ms}$, followed by lower-amplitude oscillations of similar period, as seen in the total waveform (best seen in the upper row of Figure~\ref{fig:decomposition}.C, corresponding to the model with the largest rotational rate).

\subsubsection{\label{sec:qualitative_PNSvibe}PNS core vibrations}

  After bounce, the \gls{GW} signal contains two identifiable components: one from oscillations of the \gls{PNS} core and another from convection in the sonic envelope. These regions are not fully decoupled; they interact via hydrodynamic waves and convective penetration across their interface—often termed overshoot (outwards) or undershoot (inwards).
  In this section we introduce the contribution from \gls{PNS}-core vibrations and discuss convection in Section~\ref{sec:qualitative_proco}.

As was noted in the previous section, all models exhibit prompt post-bounce core vibrations. Their aspherical components excite a high-frequency \gls{GW} signal ($f \gtrsim \unit[500]{Hz}$) that originates near the centre and propagates outwards (2, 3). These vibrations propagate with different frequencies before crossing the \gls{PNS} core. High-frequency oscillations ($f \gtrsim \unit[1000]{Hz}$) originate near the centre and propagate to $r \sim \unit[10]{km}$ (2) before merging into lower-frequency components ($f \sim \unit[750]{Hz}$, 3), which then continue until they cross the \gls{PNS} core. Upon crossing, these modes lose most of their energy and decrease their intensity, propagation velocity, and frequency ($f \leq \unit[400]{Hz}$, 4).

  In the non-rotating model, the post-bounce configuration is nearly spherical, emitting \glspl{GW} with negligible amplitude. Asymmetries are seeded by small deformations of the shock advected down to the core (5). After about $\unit[7]{ms}$, non-spherical oscillation modes develop, and \gls{GW} emission at $f \gtrsim \unit[500]{Hz}$ sets in. The oscillations, and corresponding \gls{GW} signals, are stronger between $t \approx \unit[15]{ms}$ and $t \approx \unit[30]{ms}$, but continue with generally lower amplitude for another $\approx \unit[40]{ms}$.
  In contrast, the \gls{PNS} cores in rotating models are aspherical already when they form (see region (1) in Figure~\ref{fig:decomposition}.B,C). Their anisotropic oscillations emit ringdown modes immediately after bounce.

  The intermediate-rotation model maintains noticeable oscillations throughout the entire $\unit[100]{ms}$ interval displayed in Figure~\ref{fig:decomposition}.B, though with reduced intensity near $\unit[50]{ms}$ and $\unit[100]{ms}$. After an initial stabilisation, rotation re-excites the core around $\unit[60]{ms}$ post-bounce. The resulting oscillations propagate outwards at the sound speed ($c_\textnormal{s} \sim 4$–$9\cdot10^9\,\textnormal{cm}\,\textnormal{s}^{-1}$). They experience slight damping  when entering the small convectively unstable region outside the core (shaded around $r \approx \unit[20]{km}$), but reappear clearly farther out as vertical stripes spanning both core and sonic envelope. This re-excitation occurs when the epicyclic frequency (or one of its overtones) of fluid layers around the core matches a specific mode, leading to resonance \citep[see][]{Cusinato25}. In this case, a mode near $\unit[750]{Hz}$ is excited (6).  

  The fast-rotation model does not show renewed activity after $t \sim \unit[40]{ms}$. The absence of re-excitation can be explained by rotational suppression of core deformations, consistent with earlier studies \citep{Dimmelmeier_2008PhRvD..78f4056,Shibata_2005PhRvD..71b4014}.

  In all cases, the core strain retains a high characteristic frequency. Its amplitude decreases in all models except the intermediate-rotation case, where resonance amplifies it \citep{Cusinato25}, consistent with the findings of \cite{Westernacher-Schneider2019}.

\subsubsection{\label{sec:qualitative_proco}Convection}

  Following bounce, negative entropy and $Y_e$ gradients develop in the sonic envelope, triggering prompt convection in both non-rotating \citep{Bruenn94} and rotating models. This phase lasts until about $\unit[25]{ms}$, during which the convective region spans most of the sonic envelope and extends below the \gls{PNS} core boundary. The instability criterion, $N^2 < 0$, shown in Figure~\ref{fig:decomposition} only approximates this behaviour: convection begins where the squared Brunt–Väisälä frequency becomes negative (the shaded region), but quickly 
  overshoots inwards, nearly reaching the core.

  The resulting strong, chaotic fluid motions last up to $\sim\unit[40]{ms}$ and produce positive and negative $A_{20}^{E2}$ with large amplitude in the sonic envelope (7). Eventually, the radial extent of the convective layer shrinks. Typical frequencies are $f \lesssim \unit[500]{Hz}$. As prompt convection subsides, activity retreats to the outer edge of the sonic envelope, producing weak advection modes that propagate inwards (8).

   During this phase, the sonic-envelope strain dominates the overall emission, with a pronounced peak around $10-\unit[20]{ms}$ followed by weaker activity. Its intensity is comparable in fast- and slow-rotating models, but enhanced in the intermediate case.

  Throughout the evolution, the outer-region strain remains comparatively weak, reflecting mainly the overshooting motions of prompt convection above the sonic envelope and centrifugal effects on infalling matter. Slower equatorial infall (relative to the axis) produces a quadrupole mass moment and thus weak emission. This effect grows with rotation rate.

\subsection{\label{sec:EEMD_Dec}GW ensemble empirical mode decomposition}

\begin{figure}[t]
    \centering
    \resizebox{\hsize}{!}{\includegraphics{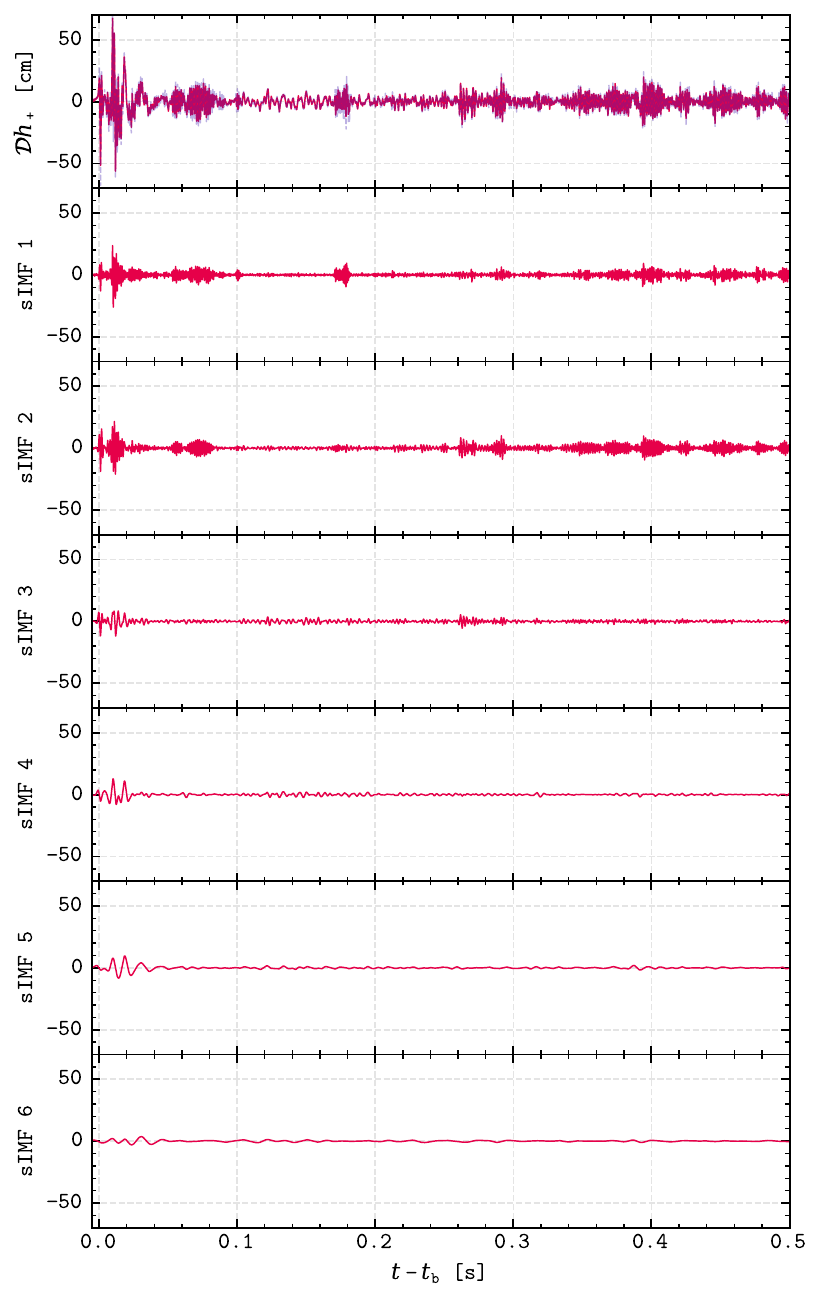}}
    \caption{Time evolution of the total \gls{GW} strain and the six \acrshortpl{sIMF} for model \texttt{s1.2-1}. The dashed blue line on the first panel represents the full strain extracted  on the whole simulation domain, and the solid red line the strain as a sum of the \acrshortpl{sIMF}.}
    \label{fig:EEMD_decomposition}
\end{figure}

After removing the residuals (see point (iii)  of Section \ref{sec:EEMD}), we decomposed our filtered waveforms into 10 \glspl{IMF} as is described in Section \ref{sec:EEMD}. We used the matching score, $M$, as a metric to determine which of the \glspl{IMF} contribute to the overall waveform \citep{Suvorova19}:
\begin{equation}
    \label{eq:matching}
    M = \frac{(h|\tilde{h})}{\sqrt{(h|h)(\tilde{h}|\tilde{h})}},
\end{equation}
where $h$ is the original signal with the \gls{EMD} residual removed, $\tilde{h}$ is the signal obtained by summing a certain number of \glspl{IMF}, and $(a|b)$ denotes the inner product.

By analysing the matching score, we notice that the first two and the last two \glspl{IMF} contribute negligibly to the strain.
The former are associated with high-frequency numerical noise, while the latter are too weak to be significant components of the original signal. Hereafter, we identify the first \gls{sIMF} as the third one produced by the process (see Figure \ref{fig:EEMD_decomposition} for the signal decomposition into \glspl{sIMF}). Moreover, the matching coefficient exceeds 0.95 in every case. Therefore, we conclude that the overall \gls{GW} strain is consistently approximated by the first six \glspl{sIMF}.

In the case of jet formation, the outflow of ejected matter causes the \gls{GW} signal to drift to positive values (see e.g.~ the case of models \texttt{s1.0-4} and \texttt{s1.8-4} in Figure \ref{fig:GWs_set}). This drift translates into a non-periodic signal enclosed in the residual of the \gls{EMD}. For this reason, we remove the residual from the waveform to obtain a signal that oscillates around zero.

Following the qualitative observations made in Section \ref{sec:qualitative}, and particularly the observation on the characteristic frequencies of the strain in each of the analysis regions, we find that the core strain is well approximated by \glspl{sIMF} 1, 2, and 3, while the convective part of the strain corresponds to \glspl{sIMF} 4, 5, and 6. Using the matching coefficient to probe the alignment between this prescription and the strain extracted from the hydrodynamic quantities, we find that it is above 0.6. This lower matching score is likely a result of the fact that modes are not strictly confined to the regions where they originate; rather, they can propagate into or influence neighbouring regions (e.g. fast modes generated in the \gls{PNS} core and propagating through the sonic envelope). From now on, we refer to the sum of the first three \glspl{sIMF} as \gls{PNS} core strain, $h_{\textnormal{core}}$, and to the sum of the last three after the bounce time (which we conventionally take as $5\,$ms) 
as convection strain, $h_{\textnormal{conv}}$.

\subsection{Bounce and convection signals}
\label{sec:bounce_convection}

\begin{figure}[t]
    \centering
    \resizebox{\hsize}{!}{\includegraphics{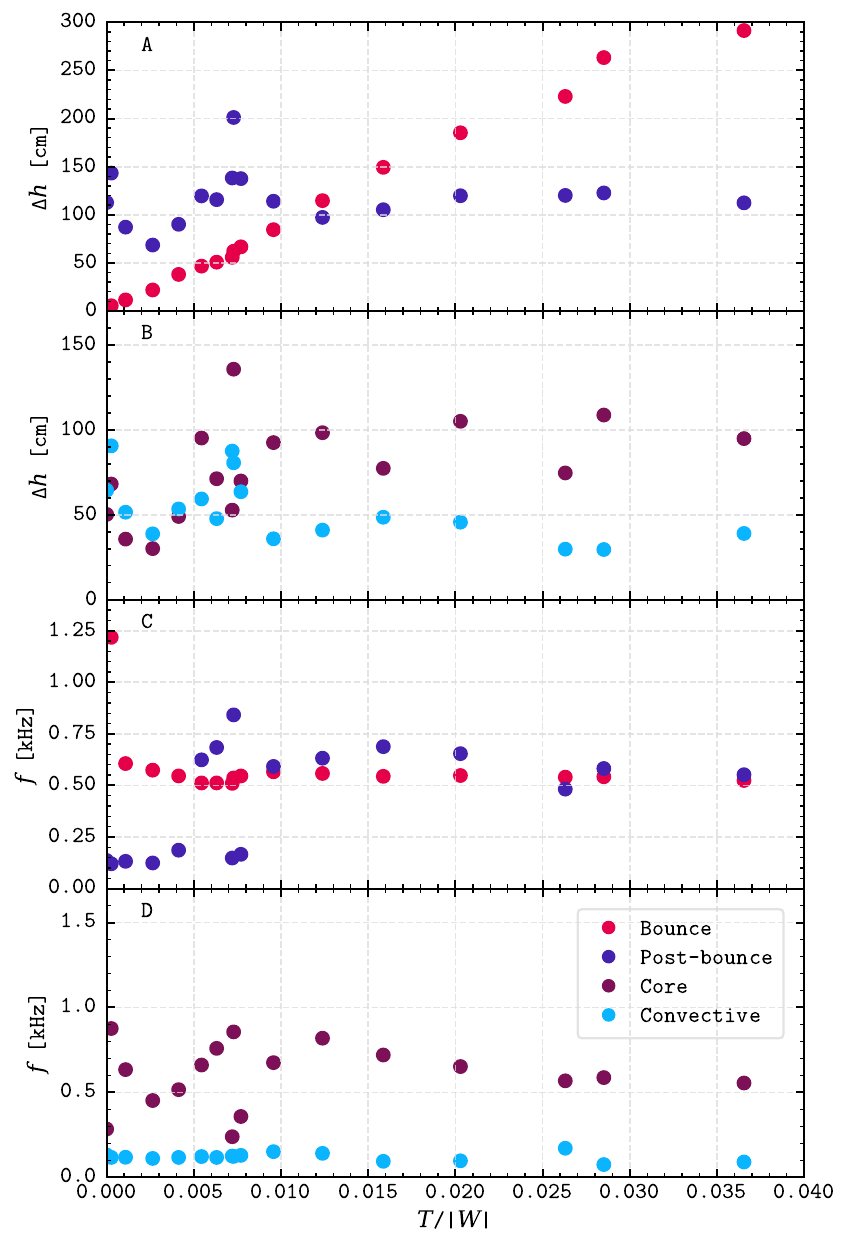}}
    \caption{Panels A and B show the difference between the highest and lowest points of the \gls{GW} signal, $\Delta h$, against the rotational kinetic energy to gravitational energy ratio for bounce and post-bounce signals, and for \gls{PNS} core~($\Delta h_{\rm core}$) and convection ($\Delta h_{\rm conv}$) signals, respectively, for models with magnetic field configuration \texttt{1}. Panels C and D display the frequencies of the previous intensity peaks in relation to the same quantity on the x axis for the same pool of models as in the previous two panels.}
    \label{fig:peaks}
\end{figure}

\begin{figure}[t]
    \centering
    \resizebox{\hsize}{!}{\includegraphics{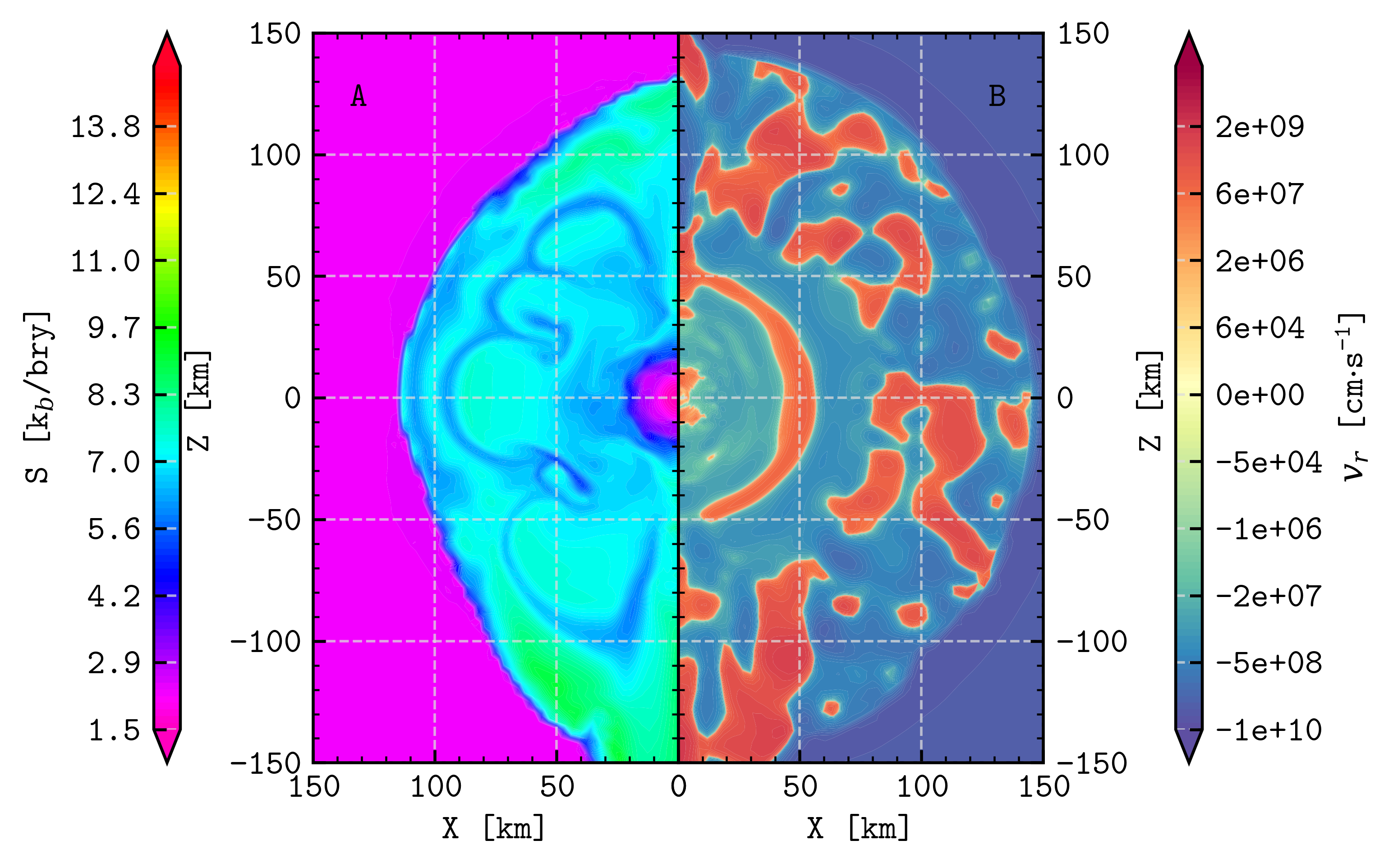}}
    \caption{Panel A: Snapshot of the entropy at \unit[15]{ms} after bounce, showing the convective bubbles surrounding the \gls{PNS} core. Panel B: Snapshot of the radial velocity at \unit[65.7]{ms}, showing the propagation (here at $\unit[50]{km}$) of a wave generated by oscillations of the \gls{PNS} core~and propagating outwards. Both snapshots are for model \texttt{s1.0-1}.}
    \label{fig:conv_bubbles}
\end{figure}

For the models with magnetic field configuration \texttt{1} (\texttt{sA.B-1} in Table \ref{tab:table_1}), the loudest signal of the early post-bounce~waveform ($t \lesssim \unit[100]{ms}$) is, depending on the rotation rate, either the bounce signal or the strong peak due to convection that appears within the following $\unit[20]{ms}$. 
We compute the difference between the highest and lowest points in the bounce signal ($t\leq\unit[5]{ms}$), $\Delta h_{\textnormal{b}}$, and the post-bounce signal ($\unit[5]{ms}\leq t\leq\unit[100]{ms}$), $\Delta h_{\textnormal{pb}}$.

Figure \ref{fig:peaks}.A shows these differences as a function of $T/|W|$, at core bounce. As is described in \cite{Abdikamalov14} and \cite{Richers17}, the bounce signal, at the rotation rates we are considering, depends linearly on $T/|W|$,  and the slope obtained from a linear regression of our models agrees within 5\% with the prediction of \cite{Richers17} for the SFHo \gls{EOS} and corresponding $T/|W|$ values.
On the contrary, $\Delta h_{\textnormal{pb}}$ fluctuates between $\unit[70]{cm}$ and $\unit[200]{cm}$ as $T/|W|$ increases, with its lowest and highest values occurring when $T/|W|$ is 0.003 and 0.007, respectively. Additionally, we observe that for slow and intermediate rotational rates, the magnitude of the post-bounce signal varies more significantly and also dominates over the magnitude of the bounce signal. In contrast, for fast rotation regimes, the magnitude of the post-bounce signal remains relatively unchanged across different values of $T/|W|$.

As is observed in Section \ref{sec:qualitative}, the post-bounce signal comprises different modes oscillating at various frequencies. Therefore, the oscillations in magnitude can be attributed to their constructive or destructive interference. To better understand this, in Figure \ref{fig:peaks}.B we separately study the intensity of the signals arising from the convection, $\Delta h_\textnormal{conv}$, and \gls{PNS} core, $\Delta h_\textnormal{core}$, as sums of their corresponding \glspl{sIMF} (see Section \ref{sec:EEMD_Dec}).
For the entirety of our sample, the convection signal reaches the maximum strain width around $\unit[15-20]{ms}$ (Figure \ref{fig:peaks}.B), which indicates that prompt post-bounce convection causes the strongest \gls{GW} signal in the convective strain.
After the bounce, gradients in entropy and lepton fraction form due to the weakening shock and core deleptonisation, driving the build-up of convective bubbles surrounding the \gls{PNS} core~(see Figure \ref{fig:conv_bubbles}.A).
Convective motions cause the \gls{GW} amplitude to grow but also tend to flatten the entropy and electron fraction gradients. When this happens, the \glspl{GW} amplitude growth stops. Before convection ceases, it reaches its maximum strength, marked by the peak in $\Delta h_{\rm conv}$ and its associated \gls{GW} signal.
Moreover, we see that the convection strain range width, $\Delta h_{\textnormal{conv}}$, is maximum for very slowly rotating progenitors and for a specific value of $T/|W|=0.007$, while it drops to around $\unit[40]{cm}$ for all other configurations (Figure \ref{fig:peaks}.B).

In most cases, the maximum of the strain range width of the \gls{PNS} core \gls{GW}, $\Delta h_{\textnormal{core}}$, and that of the convective strain are nearly coincident (Figure \ref{fig:peaks}.B, for $T/|W|\approx 0.007$). 
This is another manifestation of the resonance between the epicyclic frequency and the core normal oscillation modes found in \citep{Cusinato25}. 
While $\Delta h_{\rm core}$ is mild at low rotation rates, it peaks at intermediate ones, and then stabilises at around $\unit[90]{cm}$ in the faster rotating models, when it also becomes the part of the waveform with the larger amplitude. 
Notably, the strain range width of the core and convective strain resonates \citep{Cusinato25} at the same rotation rate (at around $T/|W|\approx 0.007$; Figure \ref{fig:peaks}.B).

In Figures \ref{fig:peaks}.C and \ref{fig:peaks}.D, we display the dominant frequency for the bounce ($f_\textnormal{b}$), post-bounce ($f_\textnormal{pb}$), \gls{PNS} core  ($f_\textnormal{core}$), and convective ($f_\textnormal{conv,peak}$) signals as a function of $T/|W|$ at bounce (see App.\,\ref{sec:freq_estimation}).
The peak frequency for the bounce signal shows little variation with increasing rotation (Figure \ref{fig:peaks}.C) as discussed by \cite{Richers17}.
However, $f_{\rm b}$ stabilises in our models  at  $\sim\unit[500]{Hz}$, whereas in their case this plateau is at $\sim\unit[700]{Hz}$. This discrepancy can be attributed to the different treatment of gravity or to the difference in the employed progenitor model.
On the other hand, $f_{\rm pb}$ behaves differently in three different regimes. In slowly rotating models, convection dominates the \gls{GW} signal over the \gls{PNS} ringdown oscillations and \gls{PNS} core vibrations. Hence, the dominant frequency is that of convection, i.e. $\unit[150-200]{Hz}$.
In intermediately rotating models, rotation causes the excitation of fast oscillating modes that in most cases dominate the \gls{GW} signal over convection, though some cases are convection-dominated. In models with fast and very fast rotation, the bounce is violent enough to produce strong intensity ringdown oscillations, with a frequency similar to that of the bounce signal.

Similarly to the intensity peaks, we also separately analyse the peak frequency of the convection and \gls{PNS} core signals. The convection signal peak frequency settles at $\sim\unit[150]{Hz}$ regardless of the model's pre-collapse rotation (Figure \ref{fig:peaks}.D), indicating that prompt convection is not significantly affected by it. On the contrary, the \gls{PNS} core peak frequency shows a non-monotonic dependence on the rotation rate, varying from $\unit[250]{Hz}$ to $\unit[900]{Hz}$ for most of the considered models.

Slowly rotating models with $T/|W|\sim 0$ and intermediate rotating models show the highest frequency peaks, but for two different reasons. The former show high-frequency modes excited by the strong convection formed post-bounce, while the latter's modes are excited in resonance with the rotation (see the peak of both $\Delta h_{\rm conv}$ and $\Delta h_{\rm conv}$ at around $T/|W|\approx 0.07$ in Figure \ref{fig:decomposition}.B). On the other hand, fast and very fast-rotating models exhibit only the ringdown oscillations from the bounce signal, with a decreasing trend $f_\textnormal{core}$ as rotation increases.

\subsection{Time evolution}
\label{sec:late_times}

\begin{figure*}[t]
    \centering
\includegraphics[width=17cm]{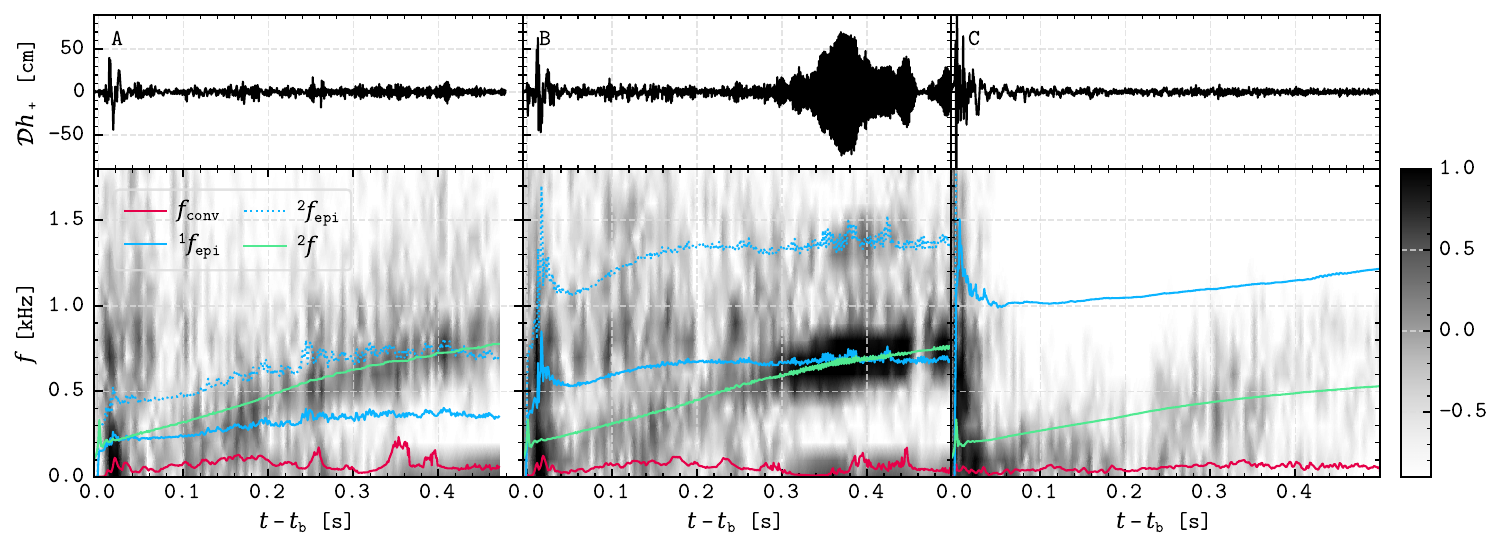}
    \caption{\gls{GW} amplitude (top row) and corresponding spectrograms for models \texttt{s0.4-1} (panel A, representative of the class of slowly rotating models), \texttt{s0.9-1} (B, prototype of intermediately rotating cases) and \texttt{s2.4-1} (C, representative of the class of fast-rotating models), calculated with a time window of  $\unit[10]{ms}$. The red line corresponds to the convection frequency (Equation (\ref{eq:conv_f})) at the sonic envelope~surface, the blue lines show the epicyclic frequency (solid) and its first overtone (dotted) in the $\pi/4$ direction at the \gls{PNS} core surface, while the green line is the fundamental quadrupolar mode frequency ($^2f$) computed with the quasi-universal relation in \cite{TorresForne19b}.}
    \label{fig:spectros}
\end{figure*}

\begin{figure}[t]
    \centering
    \resizebox{\hsize}{!}{\includegraphics{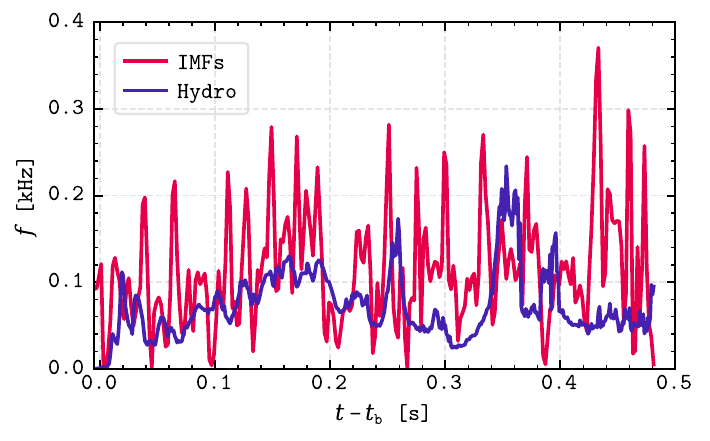}}
    \caption{Evolution of the frequency of the convection strain expressed as the \gls{IF} of the sum of the last three \glspl{sIMF} (blue line) and as the frequency associated with the convective velocity for model \texttt{s0.4-1}.}
    \label{fig:IF-conv}
\end{figure}

We considered the overall evolution of the waveform for models with the standard magnetic fields configuration (labelled as \texttt{1}), for which rotation has a major dynamic impact. Particularly, with faster rotation, the \gls{PNS} deforms, flattening at the poles and bulging at the equator due to centrifugal force. Rotation also effectively suppressing the \gls{PNS} core vibrations by dampening external excitations of \gls{PNS} vibrations  \citep{Pajkos2019}.

Figure \ref{fig:spectros} shows the \gls{GW} time evolution and corresponding spectrograms for three models (\texttt{s0.4-1}, \texttt{s0.9-1}, and \texttt{s2.4-1}), each representative of a different rotational range. 
Regardless of the initial rotation, the \glspl{GW} emission is highly active from bounce until approximately $\unit[25]{ms}$ post-bounce across a broad frequency spectrum. As described in Section \ref{sec:qualitative}, this phase is characterised by strong prompt convection, emitting both low-frequency ($f\lesssim\unit[500]{Hz}$) and high-frequency ($f\gtrsim\unit[500]{Hz}$) \glspl{GW}. A notable feature in the high-frequency category is a mode at $\unit[750]{Hz}$, which intensifies with increasing rotation rate (see Figure \ref{fig:spectros}, bottom panels).
In slow, fast, and very fast-rotating models, high-frequency modes weaken and disappear after $\sim\unit[50]{ms}$. The \gls{GW} emission then enters a quieter phase, characterised by a smaller amplitude and a steady increase in the frequency.

The behaviour of the intermediately rotating case is exceptional. The frequency of the mode excited due to rotation starts to ramp up as well (see bottom panel of Figure \ref{fig:spectros}.B). Additionally, in all rotational regimes, mode mixing renders the fundamental $^2f$ mode (\cite{TorresForne19b, Rodriguez23}, green lines) and the convection modes indistinguishable in the spectrograms. Nevertheless, a cautionary note is in order here. Since the frequency of the fundamental $^2f$ mode has been obtained from universal relations computed for non-rotating \gls{CCSN} simulations, and their applicability to fast-rotating cases is not direct, the actual evolution of the $^2f$ may be somewhat different from that presented in this paper.

At timescales longer than $\approx \unit[25]{ms}$, the low-frequency \gls{GW} signal is no longer associated with prompt convection. Instead, it originates from longer-lived convective activity in the post-shock region, which we characterise through the Brunt–Väisälä frequency (Equation~\ref{eq:BV_freq}). For clarity, we will refer to this later signal as `convection', while the early-time contribution will be explicitly denoted as `prompt convection'. It is also worth noting that, on similar timescales, additional non-convective contributions such as \gls{SASI} may also produce low-frequency \gls{GW} emission (see e.g, \citep{Kuroda16, Andresen17,Mezzacappa2023}.

For slowly and intermediately rotating models, the broadband interval of frequencies associated with the $^2f$ mode gradually separate from the convection mode between $\unit[0.1-0.2]{s}$, reaching frequencies between $\unit[700-1000]{Hz}$ at $\unit[0.5]{s}$ post-bounce (bottom panels of Figures \ref{fig:spectros}.A and \ref{fig:spectros}.B).
On the contrary, in fast- and very fast-rotating models, the separation between convection and normal modes is delayed in the spectrogram (see Figure \ref{fig:spectros}.C), beginning around $t\geq\unit[0.2-0.3]{s}$ post-bounce, and even then the distinction remains less pronounced than in slower rotations (bottom panel of Figure \ref{fig:spectros}.C).
This delay can be attributed to strong centrifugal forces, which, in very fast-rotating axisymmetric models, slow down the \gls{PNS} contraction and oscillations. 
Afterwards, the frequency broadens, spanning lower ranges ($\unit[0-700]{Hz}$) but with reduced intensity. It should be noted that the universal relations predicting the $^2f$ mode frequency were derived from non-rotating configurations \citep{TorresForne19b}. Therefore, some discrepancies between the predicted and observed $^2f$-mode frequencies may arise in the most rapidly rotating configurations.

After the crossing with the $^2f$ mode, convection modes maintain a steady frequency at $\sim \unit[100]{Hz}$. 
In slow and intermediate rotation cases, these modes tend to strengthen beyond $\sim\unit[0.3]{s}$, while in fast and very fast rotations they diminish over time. 

In intermediately rotating models, rotation resonates with specific \gls{PNS} modes, exciting them \citep{Cusinato25}. To characterise the frequency of fluid element oscillations in the cylindrical radial direction within our rotating \gls{PNS}, we use the epicyclic oscillation frequency:
\begin{equation}
    \label{eq:epicyclic_frequency_overtone}
    ^nf_\textnormal{epi} = n \frac{\kappa}{2\pi}, 
\end{equation}
where $n$ is the overtone number and $\kappa$ is the epicyclic frequency, defined as
\begin{equation}
    \label{eq:epicyclic_frequency}
    \kappa^2 = \frac{2\Omega}{R} \frac{\textnormal{d}(\Omega R^2)}{\textnormal{d}R},
\end{equation}
with $R$ being the cylindrical radius. Specifically, we employed the average of the local maxima of this quantity in a conical region spanning $[15^\circ,\,85^\circ]$ inside of the \gls{PNS} core. 

In the intermediately rotating models, a mode at $\unit[750]{Hz}$, excited during bounce, resonates with the fundamental epicyclic frequency (solid blue line in the lower panel of Figure \ref{fig:spectros}.B), producing multiple \gls{GW} bursts of varying amplitudes, of which the most noticeable appears at $\unit[0.30-0.45]{s}$ (see also e.g. the first two rows of Figure \ref{fig:decomposition}.B at $\unit[0.06-0.09]{s}$). These bursts result from a resonance between \gls{PNS} core oscillations and the rotational frequency (see Figure \ref{fig:conv_bubbles}.B).
When the fundamental epicyclic frequency crosses the $^2f$ mode, it resonates with it, leading to an increase in \gls{GW} amplitude to values comparable to $\Delta h_\textnormal{pb}$ around $\sim\unit[0.3]{s}$ that lasts until the end of the simulation (top panel of Figure \ref{fig:spectros}.B). Additionally, soon after this time, the first overtone of the epicyclic frequency excites a mode at $\unit[1500]{Hz}$.
The energy carried by \glspl{GW} decreases as rotation increases. However, in intermediate rotation models, the resonance between the \gls{PNS} modes and rotation breaks this trend and leads to a higher energy emission (see e.g. $E_\textnormal{GW,300}$ and $E_\textnormal{GW}$ columns in Table \ref{tab:results}).
In all models, the dominant mode remains the fundamental $^2f$ mode regardless of the initial rotation.

Using Equation (\ref{eq:matching}),  we compare the $^2f$-mode characteristic frequency obtained with the universal relations in \cite{TorresForne19b} with the \gls{IF} from the sum of the first three \glspl{sIMF}. Our models consistently yield a matching score above 0.91, confirming that the rotation rates considered here do not induce significant deviations.

Finally, Figure \ref{fig:IF-conv} compares the \gls{IF} derived from the convection strain and $f_\textnormal{conv}$, calculated as the mean value of Equation (\ref{eq:conv_f}) within a region extending from $15^\circ$ to $165^\circ$ in colatitude and radially form $\unit[30]{km}$ above to $\unit[30]{km}$ below the sonic envelope boundary for model \texttt{s0.4-1}. This angular range avoids axis-related interference, while the radial bounds have been chosen to capture the full effect of convection.
Using Equation (\ref{eq:matching}) for both frequencies yields a matching score above 0.7 for all models, except for slow and intermediate rotating strongly magnetised rotating models (\texttt{s0.6-4} and \texttt{s1.0-4}) which have a score of 0.45 and 0.5, respectively.

\subsection{Effect of magnetic fields}
\label{sec:B_field}

\begin{figure}[t]
    \centering
    \resizebox{\hsize}{!}{\includegraphics{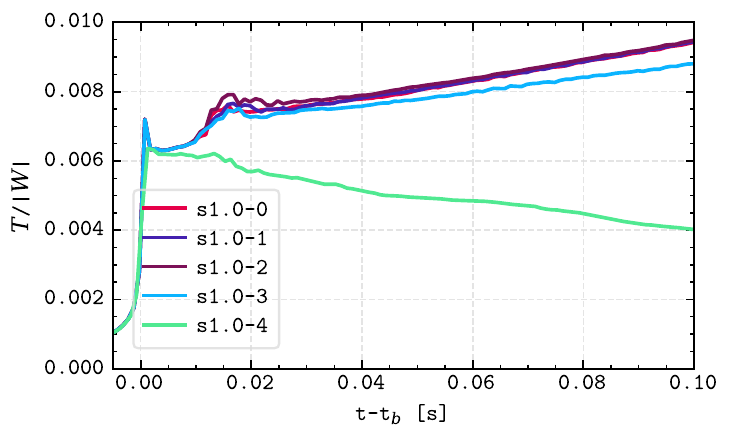}}
    \caption{Evolution of the first $\unit[100]{ms}$ of the total rotational kinetic energy of \gls{PNS} core and sonic envelope divided by the total gravitational energy for the four magnetic field configurations of model \texttt{s1.0}. }
    \label{fig:T_W_evo}
\end{figure}

\begin{figure*}[t]
    \centering
    \includegraphics[width=17cm]{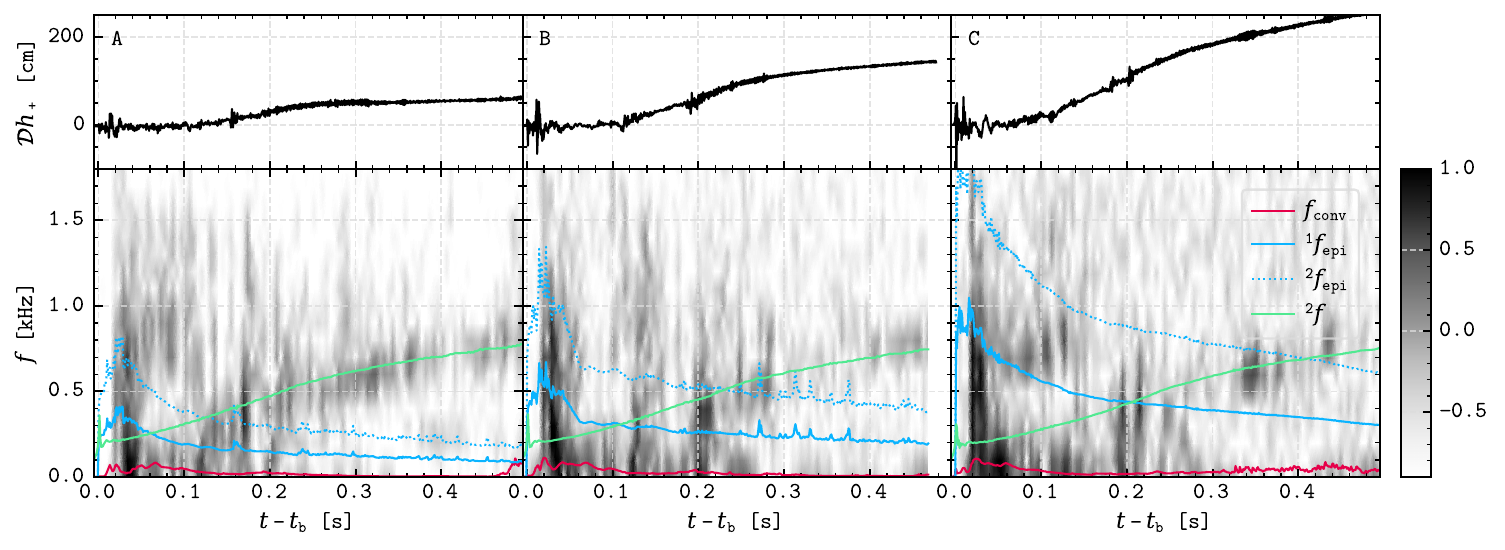}
    \caption{Same as Figure~\ref{fig:spectros} but for models \texttt{s0.6-4}, \texttt{s1.0-4}, and \texttt{s1.8-4}. The residual was removed to prevent the low-frequency memory signal from masking the convection signal.}
    \label{fig:spectro_B}
\end{figure*}

For three of the four regimes that we identified -- slow rotation, intermediate rotation, and fast rotation -- in addition to the standard pre-collapse magnetic field configuration (labelled as \texttt{1}), we ran four more models with different initial magnetic field configurations. The additional configurations are non-magnetised (\texttt{0}), or possess ten (\texttt{2}), a hundred (\texttt{3}), and a thousand times (\texttt{4}) the initial field strength.

\paragraph{Bounce and convection signals.} The magnetic field configuration, particularly at higher strengths, significantly affects the evolution of $T/|W|$. Figure \ref{fig:T_W_evo} shows the time evolution of $T/|W|$ during the first $\unit[100]{ms}$ for the four configurations of model \texttt{s1.0}. 
While $T/|W|$ peaks at bounce time in all cases, configurations \texttt{0} through \texttt{3} exhibit similar peak values. In contrast, the peak value of configuration \texttt{4} shows fluctuations as the rotation varies. 
It decreases in models with slow and intermediate rotations but increases in the fast-rotating model compared to their respective reference configurations (see $T/|W|$ column in Table \ref{tab:results}).

The post-bounce evolution follows a consistent trend across rotation rates: $T/|W|$ for configurations \texttt{0} to \texttt{3} increases rapidly until approximately $\sim\unit[15]{ms}$. Afterwards, the rate of increase slows, with configuration \texttt{3} experiencing the slowest growth. On the contrary, configuration \texttt{4} behaves differently due to magnetic braking, which decelerates the core thus lowering $T/|W|$.

The bounce strain range width and frequency are not significantly affected by the strength of the magnetic fields (see $\Delta h_\textnormal{b}$ in Table \ref{tab:results}), as magnetic braking does not have sufficient time to influence core rotation.
However, the magnetic field affects how the post-bounce, core, and convection range widths depend on the rotation rate. In the slow and fast-rotating models, most values of $\Delta h_\textnormal{pb}$, $\Delta h_\textnormal{core}$, and $\Delta h_\textnormal{conv}$, along with their associated frequencies for all magnetic field configurations, remain close to the reference values (see Table \ref{tab:results}). However, configurations \texttt{0} and \texttt{3} for model \texttt{s0.6} show a stronger strain range width. This discrepancy arises from a higher degree of asymmetry in the location of the convective bubbles forming during the prompt convection phase (e.g. similar to the ones shown in Figure \ref{fig:conv_bubbles}.B). 
For the intermediate rotation model, the range strain widths and frequencies span a broader range of values depending on the magnetic field strength. This occurs because the system is in a resonant regime, where small changes due to the magnetic fields changing the dynamics can either enhance or weaken the resonance.

\paragraph{Time evolution.} We find that in non-magnetised models (configuration \texttt{0}), the waveform evolution resembles that of models with the standard configuration, regardless of the initial rotation rate, suggesting that the magnetic fields of the standard configuration are too weak to significantly alter the dynamics.

In the following, we focus on models in which stronger fields cause major differences from the cases discussed above in Section~\ref{sec:late_times}.  Figure~\ref{fig:spectro_B} compares the time evolution and corresponding spectrograms of \glspl{GW} for magnetic field configuration \texttt{4} in models \texttt{s0.6}, \texttt{s1.0}, and \texttt{s1.8}. The spectrograms were computed by removing the residual from the original signal, so that the convection signature is not masked by the low-frequency memory signal associated with the jet structure.

In the stronger magnetised models (configurations \texttt{2}, \texttt{3}, and \texttt{4}), rotation amplifies the magnetic field along the symmetry axis, leading to jet formation. These polar jets begin forming at around $\unit[200]{km}$ from the centre, with their formation timescale largely determined by the strength of the initial magnetic field components. Moreover, mildly relativistic matter moving within the jets causes a slow shift of the \gls{GW} signal to positive values (top panels of Figure \ref{fig:spectro_B}). This memory effect becomes particularly significant when the jets are well developed, extending to approximately $\sim\unit[800]{km}$ from the star's centre, and widens progressively as the jets propagate outwards. Additionally, stronger initial magnetic fields and faster rotation rates result in a greater deviation of the \gls{GW} signal from zero.

For the intermediate-rotation case, magnetic braking in configurations \texttt{0} through \texttt{3} is sufficient to significantly slow down the core, allowing for the same resonance with the epicyclic frequency or one of its overtones observed before. However, the strong magnetic fields in configuration \texttt{4} cause substantial magnetic braking,  slowing down the core. Therefore, the epicyclic frequency no longer coincides with the \gls{PNS} core vibrations responsible for high-frequency bursts of \glspl{GW} in the post-bounce phase, resulting in the disappearance of these vibrations after $\sim\unit[50]{ms}$ (top panel of Figure \ref{fig:spectro_B}.B).

On the contrary, for the strongly rotating model \texttt{s1.8}, the magnetic braking due to the strong fields in configuration \texttt{4} slows the \gls{PNS} core sufficiently for the epicyclic frequency and its first overtone to cross the $^2f$ mode frequency at $\simeq \unit[200]{ms}$ and $\simeq \unit[420]{ms}$ after bounce, respectively (Figure \ref{fig:spectro_B}.C). These mode crossings imprint moderate-amplitude oscillations, which are much smaller than in model \texttt{s0.9-1}.

By significantly slowing down the \gls{PNS} core, the magnetic field configuration \texttt{4} reduces the energy emitted as \glspl{GW} by the intermediate-rotation model during the post-bounce phase compared to the cases with weaker fields. Among the fast-rotating models, the one with the strongest field emits the most energy (column $E_\textnormal{GW,300}$ in Table \ref{tab:results}).

Additionally, in the bottom panels of Figure \ref{fig:spectro_B}, we plot $f_\textnormal{conv}$ as approximated by Equation (\ref{eq:conv_f}). For the intermediate and slow rotating models with magnetic field configuration \texttt{4}, we observe that the frequency of the convective strain tends to be lower compared to their less magnetised counterparts. We attribute
this behaviour to strong magnetic fields hindering convection. 

We conclude this section by checking whether the usefulness of the \gls{IF} as a tool for identifying the convective dynamics is affected by the presence of magnetic fields. We find support for this assumption in the form of a matching score between the last three \glspl{sIMF} and the $f_\textnormal{conv}$ derived from the hydrodynamic variables, calculated using Equation (\ref{eq:matching}), above 0.6 for all models except for \texttt{s1.0-0} and \texttt{s0.6-4}, where the match is reduced to 0.4 and 0.5, respectively.

\section{Discussion}
\label{sec:discussion}

\subsection{Prospects of detection}
\label{sec:det_prospects}

\begin{figure*}[t]
    \centering
    \includegraphics[width=17cm]{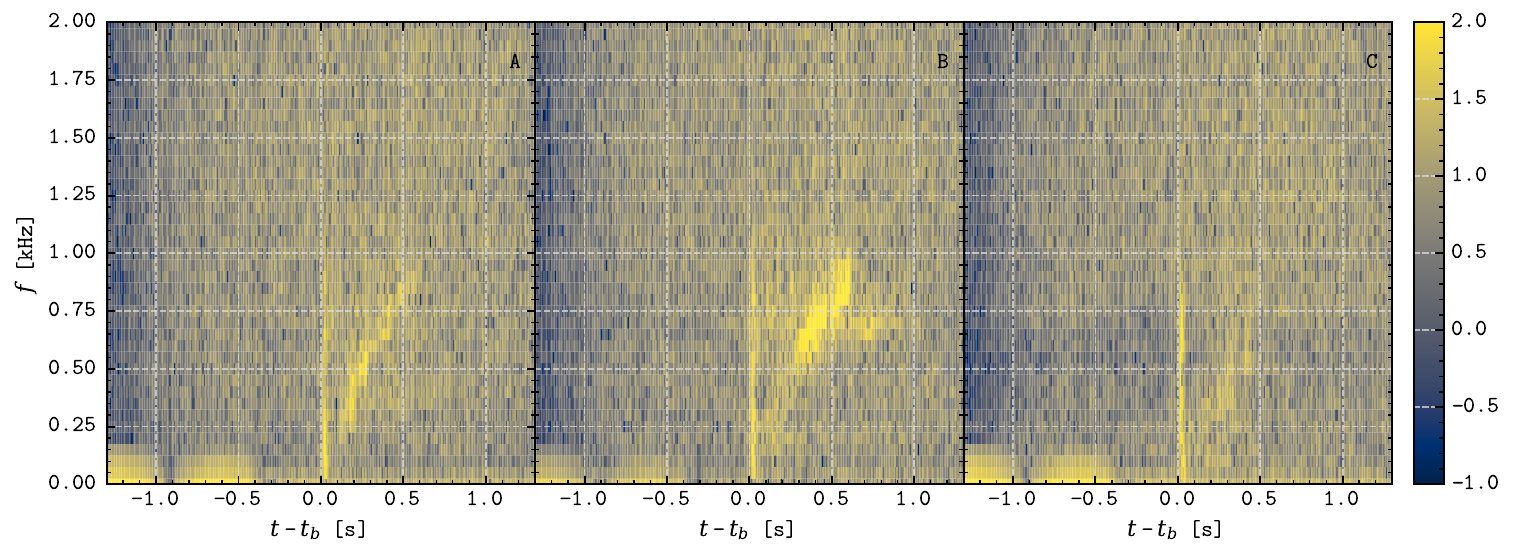}
    \caption{Time-frequency diagrams of the \gls{GW} strain at $\unit[10]{kpc}$ injected into Gaussian noise for models \texttt{s0.2-1} (panel A), \texttt{s0.9-1} (panel B) and \texttt{s1.8-1} (panel C), as seen by the LIGO Livingston interferometer for an event occurring at RA $\unit[16.99]{h}$, dec $\unit[-35.59]{^\circ}$ and arrival time $\unit[1393213818]{s}$.}
    \label{fig:spectro_noisy}
\end{figure*}

By comparing panels A, B, and C of Figure \ref{fig:spectro_noisy}, which show a slow (A), an intermediate (B) and a fast-rotating progenitor (C), we notice that a strong (broad-band and strongly time-localised) signal comes from the early post-bounce convection and ringdown in the slow and intermediate rotators, while in the fast rotator, the most prominent signal comes from the bounce. Due to the \gls{PNS} ringdown oscillations, the combination of frequencies of these two signals is very similar, posing a challenge in determining the precollapse rotation of the progenitor in the event of a \gls{GW} detection from a \gls{CCSN}. Moreover, we observe that models with higher \gls{GW} energy emitted during the bounce or post-bounce phase (see Table \ref{tab:results} and App.~\ref{sec:detectability_app} for the details of the analysis) can be observed at higher distances.

The inference of $\sqrt{M/R^3}$ is more challenging than the bounce signal due to the large variety of modes that play a role in the waveform and the much weaker signal produced. With an emitted energy of $\lesssim\unit[10^{45}]{erg}$ (see $E_{\rm GW,pb}$ in Table \ref{tab:results}), the signal emitted by very fast-rotating \glspl{CCSN} after the post-bounce peak is too weak to be detected even in events happening as close as $\unit[1]{kpc}$. The possibility of detections reported in models such as \texttt{s2.0-1} and \texttt{s2.4-1}, at higher distances, is likely a false positive of the Gaussian noise.

Progenitors with slow and fast initial rotation can be well reconstructed, at least for the best scenario, until $\unit[30]{kpc}$. Finally, most of the intermediate rotating models, even if they generally emit more energetic \glspl{GW}, due to the presence of resonant modes, can only be partially recovered. This is because, especially in the early post-bounce phase, the higher frequencies excited by resonance can interfere with the mode tracking, leading to an incomplete recovery.

If, instead of the current generation detectors, we used third-generation interferometers such as the \gls{ET} and \gls{CE}, located at the current Virgo and LIGO sites, respectively, with their theoretical sensitivity \citep{Hild11, LVK17}, the results would improve significantly: the bounce signal of all models can be recovered with at least one detector for the best scenario. 
While $\sqrt{M/R^3}$ remains hard to infer for very fast-rotating progenitors, detection for all other configurations improves, allowing for a good match until $\unit[70]{kpc}$. Moreover, the better sensitivity allows for a better resolution of the modes even in resonating cases, leading to a good inference of $\sqrt{M/R^3}$ for these models.

\subsection{Limitations of this study}
\label{sec:limitations}

This study has several limitations, which could impact the generalizability and applicability of the results.
The first one arises from considering only \gls{2D} \gls{CCSN} simulations, which produce only one \gls{GW} strain polarisation, $h_{+}$. This restriction limits our ability to fully characterise the \gls{GW} signals, as real astrophysical events occur in three dimensions (\gls{3D}) and typically generate two independent \gls{GW} polarisations. Therefore, the findings derived from these \gls{2D} simulations might not accurately represent the complexities and nuances of actual \gls{CCSN} events. Additionally, it is already well known that axisymmetric simulations tend to overestimate the amplitude of the waveform.

Moreover, our study only considered a single red supergiant progenitor model. This model was evolved through the main sequence without rotation and had a standard compactness at the pre-collapse~stage. Limiting the study to one progenitor model restricts the scope of our conclusions, as different progenitor characteristics, such as initial mass, metallicity, and compactness, can significantly influence the dynamics of the collapse and the resulting \gls{GW} signals. By not incorporating a broader range of progenitor models, we may overlook the potential variability and diversity in \gls{GW} signals that could arise from different stellar evolution pathways.

Adding a magneto-rotational profile to a non-rotating progenitor is not necessarily physically consistent.
There is a clear interplay between rotation and magnetic fields, which directly impacts the stellar structure.
However, the lack of sufficiently detailed grids of pre-supernova models covering both realistic core rotation rates and magnetic field strengths limits our ability to use fully self-consistent progenitors.
Parametrising rotation and magnetic fields as in Equations (\ref{eq:rotation_prof})and (\ref{eq:vector_potential})  enables us to systematically explore how each of these physical ingredients influences the gravitational wave signal.
More realistic stellar progenitors -- including rotation and magnetic fields in a self-consistent way -- would likely introduce more complex dependencies in the \gls{GW} signal, which are beyond the scope of this study.

\section{Conclusion}
\label{sec:conclusion}

We have presented 29 axisymmetric \glspl{CCSN} simulations based on a single red supergiant progenitor star, systematically exploring a range of magnetic field configurations and rotation profiles.
Our goal was to investigate the connection between \glspl{GW} and hydrodynamic convection during the early post-bounce~times.

Our analysis reveals that the \gls{GW} signal can be effectively approximated using the first six \glspl{sIMF} derived from \gls{EEMD} on the raw \gls{GW} strain.
Specifically, the first three \glspl{sIMF} reflect the contribution from the \gls{PNS} core, as well as the $^2f$ mode, while the last three are associated with the sonic envelope and its surrounding convective activity.
The large matching scores obtained in both cases reinforce the reliability of \gls{EEMD} as a tool for extracting physically meaningful information from \gls{GW} data related to the source dynamics.

For slowly and intermediately rotating models, we observed that the bounce signal is not consistently the dominant component of the waveform.
Instead, the  post-bounce~signal, typically resulting from the interference between \gls{PNS} ringdown oscillations and prompt convection, becomes more prominent.
In progenitor models with pre-collapse~rotation rates near $\unit[1]{\textnormal{rad}\cdot s^{-1}}$, both convection and \gls{PNS} core~vibrations are strongly excited, resulting in the loudest post-bounce signals seen in our simulations.

Furthermore, we confirm the results anticipated in \cite{Cusinato25} that rotation leads to an enhancement of the post-bounce~signal over longer timescales, caused by the resonant excitation of mode frequencies near the \gls{PNS} core~boundary.
This resonance boosts the \glspl{GW} energy emission, with an observed strain amplification of approximately one order of magnitude.

Across all explored configurations of rotation and magnetic fields, we consistently found that convection produces a low-frequency mode persisting throughout the evolution.
This mode is closely correlated with the frequency of convective velocity at the outer radius of the sonic envelope. We point out that the identification of this mode can be performed employing suitable \glspl{sIMF} resulting from an \gls{EEMD} of the \gls{GW} data.

Our work further highlights the impact of strong magnetic fields.
In addition to inducing jet formation and causing a \gls{GW} memory offset, strong magnetic fields also decelerate core rotation.
This deceleration either suppresses the resonance-induced modes in intermediately rotating models or triggers similar resonances in initially fast-rotating models as they spin down.

\begin{acknowledgements}

We acknowledge support from grant PID2021-127495NB-I00 funded by MCIN/AEI/10.13039/501100011033 and the European Union, as well as from the Astrophysics and High Energy Physics programme of the Generalitat Valenciana (ASFAE/2022/026), funded by MCIN and the European Union NextGenerationEU (PRTR-C17.I1), and from the Prometeo excellence programme grant CIPROM/2022/13 funded by the Generalitat Valenciana. MC acknowledges the support through the Generalitat Valenciana via the grant CIDEGENT/2019/031. MO was supported by the Ramón y Cajal programme of the Agencia Estatal de Investigación (RYC2018-024938-I). We are grateful to Tristan Bruel for providing the PNSinf pipeline.
The computations have been performed on servers Lluisvives and Tirant-4 (grant AECT-2025-2-0002) of the Servei d'Informàtica de la Universitat de València and on the Red Española de Supercomputación (RES) on MareNostrum (grants AECT-2025-1-0012 and AECT-2025-2-0006).
\end{acknowledgements}

\bibliographystyle{aa}

\begin{appendix}

\section{Notes on GW extraction}
\label{sec:GWextraction}
The \glspl{GW} in axisymmetry are derived following the formalism in Appendix C of~\cite{Obergaulinger_2006A&A...450.1107}. Particularly, we make use of their Equation (C.4) to calculate $N^{E2}_{20}$ in a region enclosed by two radii $r_1$ and $r_2$ as

\begin{equation}
\label{eq:NE220}
    \begin{split}
        N^{E2}_{20}\big|_{r_1}^{r_2} = \frac{G}{c^4}\frac{32\pi^{3/2}}{\sqrt{15}}\int^1_{-1}\textnormal{d}z\int^{r_2}_{r_1}&\frac{\textnormal{d}r^3}{3}\rho r \times\\
        &\left[v_r\left(3z^2-1\right)-3v_\theta z\sqrt{1-z^2}\right],
    \end{split}
\end{equation}
where $G$ is the gravitational constant, $c$ the speed of light, $\rho$ the density, $v_r$ the radial velocity, $v_\theta$ the polar velocity, $r$ the radius, and $z=\cos\, \theta$. 

Finally, the dimensionless strain $h$ is derived as

\begin{equation}
\label{eq:GWs_strain}
\mathcal{D}h_+\big|_{r_1}^{r_2} = -\frac{1}{8}\sqrt{\frac{15}{\pi}}\sin^2\Theta\frac{\textnormal{d}(N^{E2}_{20}\big|_{r_1}^{r_2} + N^{E2}_{20, \textnormal{corr}}\big|_{r_1}^{r_2})}{\textnormal{d}t},
\end{equation}
where $\mathcal{D}$ is the distance to the detector and $\Theta$ is the angle between the symmetry axis of the model and the line of sight towards the detector, which we assume to be $\sin^2\Theta=1$.

In post-processing, we compute the \gls{GW} strain calculating $N^{E2}_{20}$ from hydrodynamic quantities across the entire domain for each timestep, using Equation (\ref{eq:NE220}) and successively obtaining the gravitational wave strain using Equation (\ref{eq:GWs_strain}).

When computing the partial contributions to the \gls{GW} signal from the three subdomains defined in Section \ref{sec:GW_zones}, the surface terms at the subdomain boundaries are non-negligible. To account for these, we include the additional surface terms that arise from partial integration, following Equation (5) of \cite{Zha24}:
\begin{equation}
    \label{eq:NE220_corr}
    N^{E2}_{20, \textnormal{corr}}\big|_{r_1}^{r_2} = -\frac{G}{c^4}\frac{64\pi^{3}}{\sqrt{15}}\int^1_{-1}\textnormal{d}z \left(r^4 \rho v_r\right)\big|_{r_1}^{r_2}(3z^2-1),
\end{equation}
and successively use Equation (\ref{eq:GWs_strain}) to compute the gravitational wave strain.

\section{Frequency peaks of the GW strain}
\label{sec:freq_estimation}

\begin{figure}
    \centering
    \resizebox{\hsize}{!}{\includegraphics{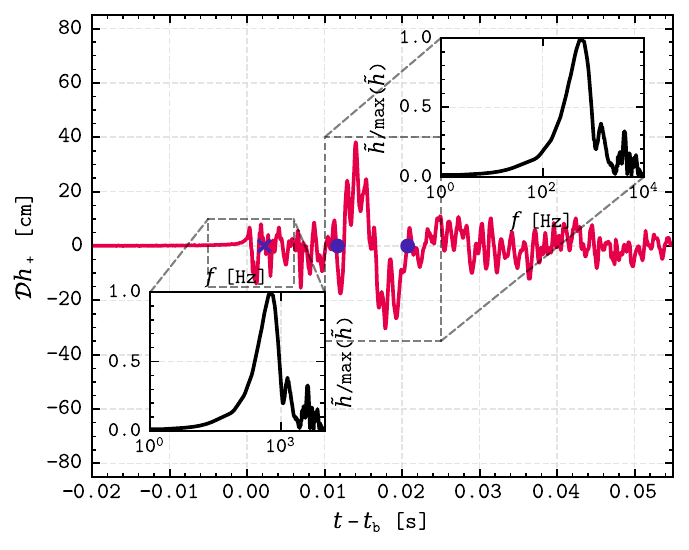}}
    \caption{\Gls{GW} evolution during the first \unit[55]{ms} for model \texttt{s0.6-1}. The blue cross marks the end of the bounce peak, and the blue circles locate the beginning and end of post-bounce oscillation. Insets show the normalised Fourier transform of the strain, $\tilde h$.}
    \label{fig:strain_points}
\end{figure}

We estimate the frequency of the peak \gls{GW} strain by taking its Fourier transform. The signal oscillates about zero, and the peak (as any other extremum of the signal) falls between two zero-crossings of the signal. As a time interval for the analysis of the frequency, we use the time across two zero-crossings before and two after the peak. Figure \ref{fig:strain_points} shows the time windows resulting from the application of the previous algorithm (maked with rectangles) over the \gls{GW} signal, along with the signal itself (red) and the normalised Fourier transforms of the strain in two insets; the lower one corresponds to analysis of the frequency around the location of $\Delta h_{\rm b}$, and the upper one to the equivalent analysis around the location of $\Delta h_{\rm pb}$. We denote with $f_{\rm b}$ and $f_{\rm pb}$ the frequencies of the spectral peak in each of the two time windows.

For the estimation of the dominant frequency associated with \gls{GW} strain of the \gls{PNS} core, $h_{\rm core}$, and of the convection, $h_{\rm conv}$, we take the Fourier transform of the corresponding strains and compute their maximum. These frequencies are denoted by $f_{\rm core}$ and $f_{\rm conv, peak}$, respectively.

\section{Detectability of computed events}
\label{sec:detectability_app}
\begin{table}
\caption{\label{tab:bounce_det} Detectability of the bounce signal for the full network of current generation detectors.}
    \centering
    \renewcommand{\arraystretch}{0.84}
    \setlength{\tabcolsep}{2.2pt}
    \begin{tabular}{>{\scriptsize}c cccccccccccccccccccc}    
        \multirow{3}{*}{{\normalsize Model}} & \multicolumn{18}{c}{Distance [kpc]} \\
& \multicolumn{2}{c}{$\unit[1]{}$} & \multicolumn{2}{c}{$\unit[5]{}$} &  \multicolumn{2}{c}{$\unit[10]{}$} & \multicolumn{2}{c}{$\unit[20]{}$} & \multicolumn{2}{c}{$\unit[30]{}$} & \multicolumn{2}{c}{$\unit[50]{}$} & \multicolumn{2}{c}{$\unit[70]{}$} & \multicolumn{2}{c}{$\unit[100]{}$} & \multicolumn{2}{c}{$\unit[200]{}$} \\        
        & B & W & B & W & B & W & B & W & B & W & B & W & B & W & B & W & B & W \\\hline
        \texttt{s0.0-1}	&	\cellcolor{cgood}	&	\cellcolor{cgood}	&	\cellcolor{cgood}	&	\cellcolor{cgood}	&	\cellcolor{cgood}	&	\cellcolor{cgood}	&	\cellcolor{cgood}	&	\cellcolor{cokish}	&	\cellcolor{cgood}	&	\cellcolor{cbad}	&	\cellcolor{cgood}	&	\cellcolor{cbad}	&	\cellcolor{cokish}	&	\cellcolor{cbad}	&	\cellcolor{cbad}	&	\cellcolor{cbad}	&	\cellcolor{cbad}	&	\cellcolor{cbad}	\\
        \texttt{s0.2-1}	&	\cellcolor{cgood}	&	\cellcolor{cgood}	&	\cellcolor{cgood}	&	\cellcolor{cgood}	&	\cellcolor{cgood}	&	\cellcolor{cgood}	&	\cellcolor{cgood}	&	\cellcolor{cokish}	&	\cellcolor{cgood}	&	\cellcolor{cbad}	&	\cellcolor{cgood}	&	\cellcolor{cbad}	&	\cellcolor{cgood}	&	\cellcolor{cbad}	&	\cellcolor{cgood}	&	\cellcolor{cbad}	&	\cellcolor{cbad}	&	\cellcolor{cbad}	\\
        \texttt{s0.4-1}	&	\cellcolor{cgood}	&	\cellcolor{cgood}	&	\cellcolor{cgood}	&	\cellcolor{cgood}	&	\cellcolor{cgood}	&	\cellcolor{cgood}	&	\cellcolor{cgood}	&	\cellcolor{cokish}	&	\cellcolor{cgood}	&	\cellcolor{cbad}	&	\cellcolor{cbad}	&	\cellcolor{cbad}	&	\cellcolor{cbad}	&	\cellcolor{cbad}	&	\cellcolor{cbad}	&	\cellcolor{cbad}	&	\cellcolor{cbad}	&	\cellcolor{cbad}	\\
        \texttt{s0.6-0}	&	\cellcolor{cgood}	&	\cellcolor{cgood}	&	\cellcolor{cgood}	&	\cellcolor{cgood}	&	\cellcolor{cgood}	&	\cellcolor{cgood}	&	\cellcolor{cgood}	&	\cellcolor{cokish}	&	\cellcolor{cgood}	&	\cellcolor{cbad}	&	\cellcolor{cgood}	&	\cellcolor{cbad}	&	\cellcolor{cbad}	&	\cellcolor{cbad}	&	\cellcolor{cbad}	&	\cellcolor{cbad}	&	\cellcolor{cbad}	&	\cellcolor{cbad}	\\
        \texttt{s0.6-1}	&	\cellcolor{cgood}	&	\cellcolor{cgood}	&	\cellcolor{cgood}	&	\cellcolor{cgood}	&	\cellcolor{cgood}	&	\cellcolor{cokish}	&	\cellcolor{cgood}	&	\cellcolor{cbad}	&	\cellcolor{cokish}	&	\cellcolor{cbad}	&	\cellcolor{cbad}	&	\cellcolor{cbad}	&	\cellcolor{cbad}	&	\cellcolor{cbad}	&	\cellcolor{cbad}	&	\cellcolor{cbad}	&	\cellcolor{cbad}	&	\cellcolor{cbad}	\\
        \texttt{s0.6-2}	&	\cellcolor{cokish}	&	\cellcolor{cgood}	&	\cellcolor{cokish}	&	\cellcolor{cgood}	&	\cellcolor{cokish}	&	\cellcolor{cgood}	&	\cellcolor{cokish}	&	\cellcolor{cokish}	&	\cellcolor{cbad}	&	\cellcolor{cbad}	&	\cellcolor{cbad}	&	\cellcolor{cbad}	&	\cellcolor{cbad}	&	\cellcolor{cbad}	&	\cellcolor{cbad}	&	\cellcolor{cbad}	&	\cellcolor{cbad}	&	\cellcolor{cbad}	\\
        \texttt{s0.6-3}	&	\cellcolor{cgood}	&	\cellcolor{cgood}	&	\cellcolor{cgood}	&	\cellcolor{cgood}	&	\cellcolor{cgood}	&	\cellcolor{cgood}	&	\cellcolor{cgood}	&	\cellcolor{cokish}	&	\cellcolor{cgood}	&	\cellcolor{cbad}	&	\cellcolor{cgood}	&	\cellcolor{cbad}	&	\cellcolor{cgood}	&	\cellcolor{cbad}	&	\cellcolor{cbad}	&	\cellcolor{cbad}	&	\cellcolor{cbad}	&	\cellcolor{cbad}	\\
        \texttt{s0.6-4}	&	\cellcolor{cgood}	&	\cellcolor{cgood}	&	\cellcolor{cgood}	&	\cellcolor{cgood}	&	\cellcolor{cgood}	&	\cellcolor{cgood}	&	\cellcolor{cgood}	&	\cellcolor{cokish}	&	\cellcolor{cgood}	&	\cellcolor{cbad}	&	\cellcolor{cgood}	&	\cellcolor{cbad}	&	\cellcolor{cokish}	&	\cellcolor{cokish}	&	\cellcolor{cbad}	&	\cellcolor{cokish}	&	\cellcolor{cbad}	&	\cellcolor{cbad}	\\
        \texttt{s0.8-1}	&	\cellcolor{cgood}	&	\cellcolor{cgood}	&	\cellcolor{cgood}	&	\cellcolor{cgood}	&	\cellcolor{cgood}	&	\cellcolor{cgood}	&	\cellcolor{cgood}	&	\cellcolor{cokish}	&	\cellcolor{cgood}	&	\cellcolor{cbad}	&	\cellcolor{cokish}	&	\cellcolor{cbad}	&	\cellcolor{cbad}	&	\cellcolor{cbad}	&	\cellcolor{cbad}	&	\cellcolor{cbad}	&	\cellcolor{cbad}	&	\cellcolor{cbad}	\\
        \texttt{s0.9-1}	&	\cellcolor{cgood}	&	\cellcolor{cgood}	&	\cellcolor{cgood}	&	\cellcolor{cgood}	&	\cellcolor{cgood}	&	\cellcolor{cokish}	&	\cellcolor{cgood}	&	\cellcolor{cokish}	&	\cellcolor{cgood}	&	\cellcolor{cokish}	&	\cellcolor{cgood}	&	\cellcolor{cbad}	&	\cellcolor{cbad}	&	\cellcolor{cbad}	&	\cellcolor{cbad}	&	\cellcolor{cbad}	&	\cellcolor{cbad}	&	\cellcolor{cbad}	\\
        \texttt{s0.95-1}	&	\cellcolor{cgood}	&	\cellcolor{cgood}	&	\cellcolor{cgood}	&	\cellcolor{cgood}	&	\cellcolor{cgood}	&	\cellcolor{cgood}	&	\cellcolor{cgood}	&	\cellcolor{cokish}	&	\cellcolor{cgood}	&	\cellcolor{cbad}	&	\cellcolor{cbad}	&	\cellcolor{cbad}	&	\cellcolor{cbad}	&	\cellcolor{cbad}	&	\cellcolor{cbad}	&	\cellcolor{cbad}	&	\cellcolor{cbad}	&	\cellcolor{cbad}	\\
    \texttt{s1.0-1}	&	\cellcolor{cgood}	&	\cellcolor{cgood}	&	\cellcolor{cgood}	&	\cellcolor{cgood}	&	\cellcolor{cgood}	&	\cellcolor{cokish}	&	\cellcolor{cgood}	&	\cellcolor{cbad}	&	\cellcolor{cgood}	&	\cellcolor{cbad}	&	\cellcolor{cgood}	&	\cellcolor{cbad}	&	\cellcolor{cokish}	&	\cellcolor{cbad}	&	\cellcolor{cokish}	&	\cellcolor{cbad}	&	\cellcolor{cokish}	&	\cellcolor{cbad}	\\
    \texttt{s1.0-0}	&	\cellcolor{cgood}	&	\cellcolor{cgood}	&	\cellcolor{cgood}	&	\cellcolor{cgood}	&	\cellcolor{cgood}	&	\cellcolor{cgood}	&	\cellcolor{cgood}	&	\cellcolor{cgood}	&	\cellcolor{cgood}	&	\cellcolor{cokish}	&	\cellcolor{cgood}	&	\cellcolor{cbad}	&	\cellcolor{cgood}	&	\cellcolor{cbad}	&	\cellcolor{cokish}	&	\cellcolor{cbad}	&	\cellcolor{cbad}	&	\cellcolor{cbad}	\\
    \texttt{s1.0-2}	&	\cellcolor{cgood}	&	\cellcolor{cgood}	&	\cellcolor{cgood}	&	\cellcolor{cgood}	&	\cellcolor{cgood}	&	\cellcolor{cgood}	&	\cellcolor{cgood}	&	\cellcolor{cgood}	&	\cellcolor{cgood}	&	\cellcolor{cgood}	&	\cellcolor{cgood}	&	\cellcolor{cgood}	&	\cellcolor{cgood}	&	\cellcolor{cokish}	&	\cellcolor{cgood}	&	\cellcolor{cbad}	&	\cellcolor{cbad}	&	\cellcolor{cbad}	\\
    \texttt{s1.0-3}	&	\cellcolor{cgood}	&	\cellcolor{cgood}	&	\cellcolor{cgood}	&	\cellcolor{cgood}	&	\cellcolor{cgood}	&	\cellcolor{cokish}	&	\cellcolor{cgood}	&	\cellcolor{cbad}	&	\cellcolor{cgood}	&	\cellcolor{cbad}	&	\cellcolor{cokish}	&	\cellcolor{cbad}	&	\cellcolor{cbad}	&	\cellcolor{cbad}	&	\cellcolor{cbad}	&	\cellcolor{cbad}	&	\cellcolor{cbad}	&	\cellcolor{cbad}	\\
    \texttt{s1.0-4}	&	\cellcolor{cgood}	&	\cellcolor{cgood}	&	\cellcolor{cgood}	&	\cellcolor{cgood}	&	\cellcolor{cgood}	&	\cellcolor{cgood}	&	\cellcolor{cgood}	&	\cellcolor{cgood}	&	\cellcolor{cgood}	&	\cellcolor{cgood}	&	\cellcolor{cgood}	&	\cellcolor{cokish}	&	\cellcolor{cokish}	&	\cellcolor{cokish}	&	\cellcolor{cokish}	&	\cellcolor{cokish}	&	\cellcolor{cokish}	&	\cellcolor{cbad}	\\
    \texttt{s1.05-1}	&	\cellcolor{cgood}	&	\cellcolor{cgood}	&	\cellcolor{cgood}	&	\cellcolor{cgood}	&	\cellcolor{cgood}	&	\cellcolor{cgood}	&	\cellcolor{cgood}	&	\cellcolor{cgood}	&	\cellcolor{cgood}	&	\cellcolor{cgood}	&	\cellcolor{cgood}	&	\cellcolor{cgood}	&	\cellcolor{cgood}	&	\cellcolor{cokish}	&	\cellcolor{cgood}	&	\cellcolor{cbad}	&	\cellcolor{cbad}	&	\cellcolor{cbad}	\\
    \texttt{s1.1-1}	&	\cellcolor{cgood}	&	\cellcolor{cgood}	&	\cellcolor{cgood}	&	\cellcolor{cgood}	&	\cellcolor{cgood}	&	\cellcolor{cgood}	&	\cellcolor{cgood}	&	\cellcolor{cgood}	&	\cellcolor{cgood}	&	\cellcolor{cokish}	&	\cellcolor{cgood}	&	\cellcolor{cbad}	&	\cellcolor{cokish}	&	\cellcolor{cbad}	&	\cellcolor{cbad}	&	\cellcolor{cbad}	&	\cellcolor{cbad}	&	\cellcolor{cbad}	\\
    \texttt{s1.2-1}	&	\cellcolor{cgood}	&	\cellcolor{cgood}	&	\cellcolor{cgood}	&	\cellcolor{cgood}	&	\cellcolor{cgood}	&	\cellcolor{cgood}	&	\cellcolor{cgood}	&	\cellcolor{cgood}	&	\cellcolor{cgood}	&	\cellcolor{cokish}	&	\cellcolor{cgood}	&	\cellcolor{cbad}	&	\cellcolor{cgood}	&	\cellcolor{cbad}	&	\cellcolor{cokish}	&	\cellcolor{cbad}	&	\cellcolor{cbad}	&	\cellcolor{cbad}	\\
    \texttt{s1.4-1}	&	\cellcolor{cgood}	&	\cellcolor{cgood}	&	\cellcolor{cgood}	&	\cellcolor{cgood}	&	\cellcolor{cgood}	&	\cellcolor{cgood}	&	\cellcolor{cgood}	&	\cellcolor{cgood}	&	\cellcolor{cgood}	&	\cellcolor{cokish}	&	\cellcolor{cgood}	&	\cellcolor{cbad}	&	\cellcolor{cokish}	&	\cellcolor{cbad}	&	\cellcolor{cbad}	&	\cellcolor{cbad}	&	\cellcolor{cbad}	&	\cellcolor{cbad}	\\
    \texttt{s1.6-1}	&	\cellcolor{cgood}	&	\cellcolor{cgood}	&	\cellcolor{cgood}	&	\cellcolor{cgood}	&	\cellcolor{cgood}	&	\cellcolor{cgood}	&	\cellcolor{cgood}	&	\cellcolor{cgood}	&	\cellcolor{cgood}	&	\cellcolor{cokish}	&	\cellcolor{cgood}	&	\cellcolor{cbad}	&	\cellcolor{cokish}	&	\cellcolor{cbad}	&	\cellcolor{cbad}	&	\cellcolor{cbad}	&	\cellcolor{cbad}	&	\cellcolor{cbad}	\\
    \texttt{s1.8-0}	&	\cellcolor{cgood}	&	\cellcolor{cgood}	&	\cellcolor{cgood}	&	\cellcolor{cgood}	&	\cellcolor{cgood}	&	\cellcolor{cgood}	&	\cellcolor{cgood}	&	\cellcolor{cgood}	&	\cellcolor{cgood}	&	\cellcolor{cokish}	&	\cellcolor{cgood}	&	\cellcolor{cbad}	&	\cellcolor{cgood}	&	\cellcolor{cokish}	&	\cellcolor{cokish}	&	\cellcolor{cokish}	&	\cellcolor{cbad}	&	\cellcolor{cbad}	\\
    \texttt{s1.8-1}	&	\cellcolor{cgood}	&	\cellcolor{cgood}	&	\cellcolor{cgood}	&	\cellcolor{cgood}	&	\cellcolor{cgood}	&	\cellcolor{cgood}	&	\cellcolor{cgood}	&	\cellcolor{cgood}	&	\cellcolor{cgood}	&	\cellcolor{cgood}	&	\cellcolor{cgood}	&	\cellcolor{cokish}	&	\cellcolor{cgood}	&	\cellcolor{cbad}	&	\cellcolor{cokish}	&	\cellcolor{cbad}	&	\cellcolor{cbad}	&	\cellcolor{cbad}	\\
    \texttt{s1.8-2}	&	\cellcolor{cgood}	&	\cellcolor{cgood}	&	\cellcolor{cgood}	&	\cellcolor{cgood}	&	\cellcolor{cgood}	&	\cellcolor{cgood}	&	\cellcolor{cgood}	&	\cellcolor{cgood}	&	\cellcolor{cgood}	&	\cellcolor{cokish}	&	\cellcolor{cgood}	&	\cellcolor{cbad}	&	\cellcolor{cokish}	&	\cellcolor{cbad}	&	\cellcolor{cbad}	&	\cellcolor{cbad}	&	\cellcolor{cbad}	&	\cellcolor{cbad}	\\
    \texttt{s1.8-3}	&	\cellcolor{cgood}	&	\cellcolor{cgood}	&	\cellcolor{cgood}	&	\cellcolor{cgood}	&	\cellcolor{cgood}	&	\cellcolor{cgood}	&	\cellcolor{cgood}	&	\cellcolor{cgood}	&	\cellcolor{cgood}	&	\cellcolor{cokish}	&	\cellcolor{cgood}	&	\cellcolor{cbad}	&	\cellcolor{cokish}	&	\cellcolor{cbad}	&	\cellcolor{cbad}	&	\cellcolor{cbad}	&	\cellcolor{cbad}	&	\cellcolor{cbad}	\\
    \texttt{s1.8-4}	&	\cellcolor{cgood}	&	\cellcolor{cgood}	&	\cellcolor{cgood}	&	\cellcolor{cgood}	&	\cellcolor{cgood}	&	\cellcolor{cgood}	&	\cellcolor{cgood}	&	\cellcolor{cgood}	&	\cellcolor{cgood}	&	\cellcolor{cgood}	&	\cellcolor{cgood}	&	\cellcolor{cokish}	&	\cellcolor{cgood}	&	\cellcolor{cbad}	&	\cellcolor{cgood}	&	\cellcolor{cbad}	&	\cellcolor{cokish}	&	\cellcolor{cbad}	\\
    \texttt{s2.0-1}	&	\cellcolor{cgood}	&	\cellcolor{cgood}	&	\cellcolor{cgood}	&	\cellcolor{cgood}	&	\cellcolor{cgood}	&	\cellcolor{cgood}	&	\cellcolor{cgood}	&	\cellcolor{cgood}	&	\cellcolor{cgood}	&	\cellcolor{cokish}	&	\cellcolor{cgood}	&	\cellcolor{cbad}	&	\cellcolor{cgood}	&	\cellcolor{cbad}	&	\cellcolor{cokish}	&	\cellcolor{cbad}	&	\cellcolor{cbad}	&	\cellcolor{cbad}	\\
    \texttt{s2.2-1}	&	\cellcolor{cgood}	&	\cellcolor{cgood}	&	\cellcolor{cgood}	&	\cellcolor{cgood}	&	\cellcolor{cgood}	&	\cellcolor{cgood}	&	\cellcolor{cgood}	&	\cellcolor{cgood}	&	\cellcolor{cgood}	&	\cellcolor{cgood}	&	\cellcolor{cgood}	&	\cellcolor{cokish}	&	\cellcolor{cgood}	&	\cellcolor{cbad}	&	\cellcolor{cgood}	&	\cellcolor{cbad}	&	\cellcolor{cbad}	&	\cellcolor{cbad}	\\
    \texttt{s2.4-1}	&	\cellcolor{cgood}	&	\cellcolor{cgood}	&	\cellcolor{cgood}	&	\cellcolor{cgood}	&	\cellcolor{cgood}	&	\cellcolor{cgood}	&	\cellcolor{cgood}	&	\cellcolor{cgood}	&	\cellcolor{cgood}	&	\cellcolor{cgood}	&	\cellcolor{cgood}	&	\cellcolor{cgood}	&	\cellcolor{cgood}	&	\cellcolor{cgood}	&	\cellcolor{cokish}	&	\cellcolor{cokish}	&	\cellcolor{cbad}	&	\cellcolor{cbad} \\ \hline \\
    \end{tabular}
    \tablefoot{Columns from left to right: model name, then best and worst case scenario for a source located in the direction of the Milky Way's bulk at varying distance, from $\unit[1]{kpc}$ to $\unit[200]{kpc}$. Cells coloured in green represent simultaneous detection with at least two detectors, cells in yellow with one detector, and in red no detection.}
\end{table}
\begin{table}
\caption{\label{tab:mr_det} Percentage of the $\sqrt{M/R^3}$ ratio inferred with a network consisting of current generation Livingston and Hanford interferometers.}
    \centering
    \renewcommand{\arraystretch}{0.84}
    \setlength{\tabcolsep}{2.2pt}
    \begin{tabular}{>{\scriptsize}c cccccccccccccccccccc}  
        \multirow{3}{*}{{\normalsize Model}} & \multicolumn{18}{c}{Distance [kpc]} \\
& \multicolumn{2}{c}{$\unit[1]{}$} & \multicolumn{2}{c}{$\unit[5]{}$} &  \multicolumn{2}{c}{$\unit[10]{}$} & \multicolumn{2}{c}{$\unit[20]{}$} & \multicolumn{2}{c}{$\unit[30]{}$} & \multicolumn{2}{c}{$\unit[50]{}$} & \multicolumn{2}{c}{$\unit[70]{}$} & \multicolumn{2}{c}{$\unit[100]{}$} & \multicolumn{2}{c}{$\unit[200]{}$} \\    
        & B & W & B & W & B & W & B & W & B & W & B & W & B & W & B & W & B & W \\\hline
        \texttt{s0.0-1}	&	\cellcolor{cgood}	&	\cellcolor{cgood}	&	\cellcolor{cgood}	&	\cellcolor{cgood}	&	\cellcolor{cgood}	&	\cellcolor{cbad}	&	\cellcolor{cgood}	&	\cellcolor{cbad}	&	\cellcolor{cbad}	&	\cellcolor{cbad}	&	\cellcolor{cgood}	&	\cellcolor{cbad}	&	\cellcolor{cbad}	&	\cellcolor{cbad}	&	\cellcolor{cgood}	&	\cellcolor{cbad}	&	\cellcolor{cbad}	&	\cellcolor{cbad}	\\
        \texttt{s0.2-1}	&	\cellcolor{cgood}	&	\cellcolor{cgood}	&	\cellcolor{cgood}	&	\cellcolor{cgood}	&	\cellcolor{cgood}	&	\cellcolor{cgood}	&	\cellcolor{cgood}	&	\cellcolor{cbad}	&	\cellcolor{cgood}	&	\cellcolor{cbad}	&	\cellcolor{cgood}	&	\cellcolor{cbad}	&	\cellcolor{cokish}	&	\cellcolor{cbad}	&	\cellcolor{cbad}	&	\cellcolor{cbad}	&	\cellcolor{cbad}	&	\cellcolor{cbad}	\\
        \texttt{s0.4-1}	&	\cellcolor{cgood}	&	\cellcolor{cgood}	&	\cellcolor{cgood}	&	\cellcolor{cgood}	&	\cellcolor{cbad}	&	\cellcolor{cgood}	&	\cellcolor{cokish}	&	\cellcolor{cgood}	&	\cellcolor{cokish}	&	\cellcolor{cbad}	&	\cellcolor{cgood}	&	\cellcolor{cbad}	&	\cellcolor{cbad}	&	\cellcolor{cbad}	&	\cellcolor{cokish}	&	\cellcolor{cokish}	&	\cellcolor{cbad}	&	\cellcolor{cbad}	\\
        \texttt{s0.6-1}	&	\cellcolor{cgood}	&	\cellcolor{cgood}	&	\cellcolor{cgood}	&	\cellcolor{cgood}	&	\cellcolor{cgood}	&	\cellcolor{cgood}	&	\cellcolor{cokish}	&	\cellcolor{cbad}	&	\cellcolor{cgood}	&	\cellcolor{cbad}	&	\cellcolor{cokish}	&	\cellcolor{cbad}	&	\cellcolor{cbad}	&	\cellcolor{cbad}	&	\cellcolor{cbad}	&	\cellcolor{cbad}	&	\cellcolor{cbad}	&	\cellcolor{cbad}	\\
        \texttt{s0.6-0}	&	\cellcolor{cgood}	&	\cellcolor{cgood}	&	\cellcolor{cgood}	&	\cellcolor{cgood}	&	\cellcolor{cgood}	&	\cellcolor{cgood}	&	\cellcolor{cgood}	&	\cellcolor{cbad}	&	\cellcolor{cbad}	&	\cellcolor{cbad}	&	\cellcolor{cgood}	&	\cellcolor{cbad}	&	\cellcolor{cgood}	&	\cellcolor{cbad}	&	\cellcolor{cbad}	&	\cellcolor{cbad}	&	\cellcolor{cbad}	&	\cellcolor{cbad}	\\
        \texttt{s0.6-2}	&	\cellcolor{cgood}	&	\cellcolor{cgood}	&	\cellcolor{cgood}	&	\cellcolor{cgood}	&	\cellcolor{cgood}	&	\cellcolor{cgood}	&	\cellcolor{cbad}	&	\cellcolor{cokish}	&	\cellcolor{cbad}	&	\cellcolor{cokish}	&	\cellcolor{cbad}	&	\cellcolor{cokish}	&	\cellcolor{cbad}	&	\cellcolor{cbad}	&	\cellcolor{cbad}	&	\cellcolor{cbad}	&	\cellcolor{cbad}	&	\cellcolor{cbad}	\\
        \texttt{s0.6-3}	&	\cellcolor{cbad}	&	\cellcolor{cbad}	&	\cellcolor{cbad}	&	\cellcolor{cokish}	&	\cellcolor{cbad}	&	\cellcolor{cokish}	&	\cellcolor{cbad}	&	\cellcolor{cbad}	&	\cellcolor{cbad}	&	\cellcolor{cgood}	&	\cellcolor{cbad}	&	\cellcolor{cbad}	&	\cellcolor{cokish}	&	\cellcolor{cgood}	&	\cellcolor{cbad}	&	\cellcolor{cbad}	&	\cellcolor{cokish}	&	\cellcolor{cbad}	\\
        \texttt{s0.6-4}	&	\cellcolor{cgood}	&	\cellcolor{cgood}	&	\cellcolor{cgood}	&	\cellcolor{cokish}	&	\cellcolor{cokish}	&	\cellcolor{cbad}	&	\cellcolor{cokish}	&	\cellcolor{cbad}	&	\cellcolor{cbad}	&	\cellcolor{cbad}	&	\cellcolor{cbad}	&	\cellcolor{cbad}	&	\cellcolor{cbad}	&	\cellcolor{cbad}	&	\cellcolor{cbad}	&	\cellcolor{cbad}	&	\cellcolor{cbad}	&	\cellcolor{cbad}	\\
        \texttt{s0.6-1}	&	\cellcolor{cgood}	&	\cellcolor{cgood}	&	\cellcolor{cgood}	&	\cellcolor{cgood}	&	\cellcolor{cgood}	&	\cellcolor{cgood}	&	\cellcolor{cgood}	&	\cellcolor{cbad}	&	\cellcolor{cbad}	&	\cellcolor{cgood}	&	\cellcolor{cokish}	&	\cellcolor{cbad}	&	\cellcolor{cokish}	&	\cellcolor{cbad}	&	\cellcolor{cbad}	&	\cellcolor{cbad}	&	\cellcolor{cokish}	&	\cellcolor{cokish}	\\
        \texttt{s0.9-1}	&	\cellcolor{cgood}	&	\cellcolor{cgood}	&	\cellcolor{cgood}	&	\cellcolor{cokish}	&	\cellcolor{cgood}	&	\cellcolor{cgood}	&	\cellcolor{cgood}	&	\cellcolor{cgood}	&	\cellcolor{cokish}	&	\cellcolor{cgood}	&	\cellcolor{cokish}	&	\cellcolor{cokish}	&	\cellcolor{cgood}	&	\cellcolor{cgood}	&	\cellcolor{cgood}	&	\cellcolor{cbad}	&	\cellcolor{cgood}	&	\cellcolor{cbad}	\\
        \texttt{s0.95-1}	&	\cellcolor{cokish}	&	\cellcolor{cokish}	&	\cellcolor{cokish}	&	\cellcolor{cokish}	&	\cellcolor{cokish}	&	\cellcolor{cokish}	&	\cellcolor{cokish}	&	\cellcolor{cokish}	&	\cellcolor{cokish}	&	\cellcolor{cokish}	&	\cellcolor{cokish}	&	\cellcolor{cokish}	&	\cellcolor{cokish}	&	\cellcolor{cokish}	&	\cellcolor{cgood}	&	\cellcolor{cbad}	&	\cellcolor{cgood}	&	\cellcolor{cgood}	\\
        \texttt{s1.0-0}	&	\cellcolor{cbad}	&	\cellcolor{cbad}	&	\cellcolor{cbad}	&	\cellcolor{cbad}	&	\cellcolor{cbad}	&	\cellcolor{cbad}	&	\cellcolor{cokish}	&	\cellcolor{cbad}	&	\cellcolor{cgood}	&	\cellcolor{cokish}	&	\cellcolor{cbad}	&	\cellcolor{cbad}	&	\cellcolor{cbad}	&	\cellcolor{cbad}	&	\cellcolor{cbad}	&	\cellcolor{cbad}	&	\cellcolor{cbad}	&	\cellcolor{cbad}	\\
        \texttt{s1.0-1}	&	\cellcolor{cgood}	&	\cellcolor{cokish}	&	\cellcolor{cokish}	&	\cellcolor{cokish}	&	\cellcolor{cokish}	&	\cellcolor{cokish}	&	\cellcolor{cokish}	&	\cellcolor{cokish}	&	\cellcolor{cokish}	&	\cellcolor{cgood}	&	\cellcolor{cokish}	&	\cellcolor{cokish}	&	\cellcolor{cokish}	&	\cellcolor{cokish}	&	\cellcolor{cokish}	&	\cellcolor{cokish}	&	\cellcolor{cokish}	&	\cellcolor{cgood}	\\
        \texttt{s1.0-2}	&	\cellcolor{cgood}	&	\cellcolor{cokish}	&	\cellcolor{cgood}	&	\cellcolor{cokish}	&	\cellcolor{cokish}	&	\cellcolor{cokish}	&	\cellcolor{cokish}	&	\cellcolor{cgood}	&	\cellcolor{cokish}	&	\cellcolor{cgood}	&	\cellcolor{cokish}	&	\cellcolor{cokish}	&	\cellcolor{cokish}	&	\cellcolor{cokish}	&	\cellcolor{cokish}	&	\cellcolor{cbad}	&	\cellcolor{cokish}	&	\cellcolor{cokish}	\\
        \texttt{s1.0-3}	&	\cellcolor{cbad}	&	\cellcolor{cbad}	&	\cellcolor{cbad}	&	\cellcolor{cbad}	&	\cellcolor{cbad}	&	\cellcolor{cbad}	&	\cellcolor{cbad}	&	\cellcolor{cbad}	&	\cellcolor{cbad}	&	\cellcolor{cbad}	&	\cellcolor{cgood}	&	\cellcolor{cbad}	&	\cellcolor{cbad}	&	\cellcolor{cokish}	&	\cellcolor{cbad}	&	\cellcolor{cbad}	&	\cellcolor{cokish}	&	\cellcolor{cbad}	\\
        \texttt{s1.0-4}	&	\cellcolor{cbad}	&	\cellcolor{cbad}	&	\cellcolor{cbad}	&	\cellcolor{cbad}	&	\cellcolor{cbad}	&	\cellcolor{cbad}	&	\cellcolor{cbad}	&	\cellcolor{cbad}	&	\cellcolor{cbad}	&	\cellcolor{cbad}	&	\cellcolor{cbad}	&	\cellcolor{cbad}	&	\cellcolor{cgood}	&	\cellcolor{cgood}	&	\cellcolor{cbad}	&	\cellcolor{cokish}	&	\cellcolor{cbad}	&	\cellcolor{cgood}	\\
        \texttt{s1.05-1}	&	\cellcolor{cgood}	&	\cellcolor{cokish}	&	\cellcolor{cgood}	&	\cellcolor{cgood}	&	\cellcolor{cgood}	&	\cellcolor{cokish}	&	\cellcolor{cgood}	&	\cellcolor{cgood}	&	\cellcolor{cgood}	&	\cellcolor{cokish}	&	\cellcolor{cgood}	&	\cellcolor{cbad}	&	\cellcolor{cokish}	&	\cellcolor{cokish}	&	\cellcolor{cgood}	&	\cellcolor{cbad}	&	\cellcolor{cbad}	&	\cellcolor{cokish}	\\
        \texttt{s1.1-1}	&	\cellcolor{cbad}	&	\cellcolor{cbad}	&	\cellcolor{cgood}	&	\cellcolor{cokish}	&	\cellcolor{cbad}	&	\cellcolor{cokish}	&	\cellcolor{cokish}	&	\cellcolor{cokish}	&	\cellcolor{cgood}	&	\cellcolor{cbad}	&	\cellcolor{cbad}	&	\cellcolor{cokish}	&	\cellcolor{cbad}	&	\cellcolor{cbad}	&	\cellcolor{cokish}	&	\cellcolor{cbad}	&	\cellcolor{cbad}	&	\cellcolor{cbad}	\\
        \texttt{s1.2-1}	&	\cellcolor{cgood}	&	\cellcolor{cgood}	&	\cellcolor{cgood}	&	\cellcolor{cgood}	&	\cellcolor{cgood}	&	\cellcolor{cgood}	&	\cellcolor{cgood}	&	\cellcolor{cgood}	&	\cellcolor{cgood}	&	\cellcolor{cokish}	&	\cellcolor{cokish}	&	\cellcolor{cbad}	&	\cellcolor{cokish}	&	\cellcolor{cbad}	&	\cellcolor{cokish}	&	\cellcolor{cokish}	&	\cellcolor{cbad}	&	\cellcolor{cbad}	\\
        \texttt{s1.4-1}	&	\cellcolor{cgood}	&	\cellcolor{cgood}	&	\cellcolor{cgood}	&	\cellcolor{cokish}	&	\cellcolor{cgood}	&	\cellcolor{cgood}	&	\cellcolor{cbad}	&	\cellcolor{cgood}	&	\cellcolor{cgood}	&	\cellcolor{cgood}	&	\cellcolor{cbad}	&	\cellcolor{cbad}	&	\cellcolor{cokish}	&	\cellcolor{cbad}	&	\cellcolor{cbad}	&	\cellcolor{cbad}	&	\cellcolor{cbad}	&	\cellcolor{cbad}	\\
        \texttt{s1.6-1}	&	\cellcolor{cbad}	&	\cellcolor{cokish}	&	\cellcolor{cokish}	&	\cellcolor{cgood}	&	\cellcolor{cbad}	&	\cellcolor{cbad}	&	\cellcolor{cbad}	&	\cellcolor{cbad}	&	\cellcolor{cokish}	&	\cellcolor{cbad}	&	\cellcolor{cbad}	&	\cellcolor{cbad}	&	\cellcolor{cokish}	&	\cellcolor{cbad}	&	\cellcolor{cbad}	&	\cellcolor{cbad}	&	\cellcolor{cbad}	&	\cellcolor{cbad}	\\
        \texttt{s1.8-1}	&	\cellcolor{cbad}	&	\cellcolor{cbad}	&	\cellcolor{cbad}	&	\cellcolor{cbad}	&	\cellcolor{cbad}	&	\cellcolor{cbad}	&	\cellcolor{cbad}	&	\cellcolor{cokish}	&	\cellcolor{cgood}	&	\cellcolor{cbad}	&	\cellcolor{cgood}	&	\cellcolor{cokish}	&	\cellcolor{cbad}	&	\cellcolor{cbad}	&	\cellcolor{cbad}	&	\cellcolor{cokish}	&	\cellcolor{cbad}	&	\cellcolor{cbad}	\\
        \texttt{s1.8-0}	&	\cellcolor{cbad}	&	\cellcolor{cbad}	&	\cellcolor{cbad}	&	\cellcolor{cgood}	&	\cellcolor{cbad}	&	\cellcolor{cbad}	&	\cellcolor{cbad}	&	\cellcolor{cbad}	&	\cellcolor{cokish}	&	\cellcolor{cgood}	&	\cellcolor{cokish}	&	\cellcolor{cokish}	&	\cellcolor{cokish}	&	\cellcolor{cbad}	&	\cellcolor{cbad}	&	\cellcolor{cokish}	&	\cellcolor{cbad}	&	\cellcolor{cokish}	\\
        \texttt{s1.8-2}	&	\cellcolor{cokish}	&	\cellcolor{cbad}	&	\cellcolor{cbad}	&	\cellcolor{cbad}	&	\cellcolor{cbad}	&	\cellcolor{cbad}	&	\cellcolor{cbad}	&	\cellcolor{cbad}	&	\cellcolor{cgood}	&	\cellcolor{cokish}	&	\cellcolor{cgood}	&	\cellcolor{cbad}	&	\cellcolor{cbad}	&	\cellcolor{cgood}	&	\cellcolor{cbad}	&	\cellcolor{cbad}	&	\cellcolor{cokish}	&	\cellcolor{cbad}	\\
        \texttt{s1.8-3}	&	\cellcolor{cbad}	&	\cellcolor{cbad}	&	\cellcolor{cbad}	&	\cellcolor{cbad}	&	\cellcolor{cbad}	&	\cellcolor{cgood}	&	\cellcolor{cbad}	&	\cellcolor{cbad}	&	\cellcolor{cbad}	&	\cellcolor{cokish}	&	\cellcolor{cokish}	&	\cellcolor{cbad}	&	\cellcolor{cbad}	&	\cellcolor{cbad}	&	\cellcolor{cbad}	&	\cellcolor{cgood}	&	\cellcolor{cbad}	&	\cellcolor{cokish}	\\
        \texttt{s1.8-4}	&	\cellcolor{cbad}	&	\cellcolor{cbad}	&	\cellcolor{cbad}	&	\cellcolor{cbad}	&	\cellcolor{cbad}	&	\cellcolor{cbad}	&	\cellcolor{cbad}	&	\cellcolor{cbad}	&	\cellcolor{cbad}	&	\cellcolor{cbad}	&	\cellcolor{cbad}	&	\cellcolor{cbad}	&	\cellcolor{cbad}	&	\cellcolor{cbad}	&	\cellcolor{cbad}	&	\cellcolor{cbad}	&	\cellcolor{cbad}	&	\cellcolor{cbad}	\\
        \texttt{s2.0-1}	&	\cellcolor{cbad}	&	\cellcolor{cbad}	&	\cellcolor{cbad}	&	\cellcolor{cgood}	&	\cellcolor{cbad}	&	\cellcolor{cgood}	&	\cellcolor{cbad}	&	\cellcolor{cbad}	&	\cellcolor{cbad}	&	\cellcolor{cbad}	&	\cellcolor{cbad}	&	\cellcolor{cbad}	&	\cellcolor{cbad}	&	\cellcolor{cbad}	&	\cellcolor{cgood}	&	\cellcolor{cbad}	&	\cellcolor{cbad}	&	\cellcolor{cbad}	\\
        \texttt{s2.2-1}	&	\cellcolor{cbad}	&	\cellcolor{cbad}	&	\cellcolor{cbad}	&	\cellcolor{cbad}	&	\cellcolor{cbad}	&	\cellcolor{cbad}	&	\cellcolor{cbad}	&	\cellcolor{cbad}	&	\cellcolor{cbad}	&	\cellcolor{cbad}	&	\cellcolor{cbad}	&	\cellcolor{cbad}	&	\cellcolor{cbad}	&	\cellcolor{cbad}	&	\cellcolor{cbad}	&	\cellcolor{cokish}	&	\cellcolor{cbad}	&	\cellcolor{cbad}	\\
        \texttt{s2.4-1}	&	\cellcolor{cbad}	&	\cellcolor{cbad}	&	\cellcolor{cbad}	&	\cellcolor{cokish}	&	\cellcolor{cbad}	&	\cellcolor{cgood}	&	\cellcolor{cokish}	&	\cellcolor{cbad}	&	\cellcolor{cgood}	&	\cellcolor{cokish}	&	\cellcolor{cokish}	&	\cellcolor{cgood}	&	\cellcolor{cbad}	&	\cellcolor{cbad}	&	\cellcolor{cokish}	&	\cellcolor{cokish}	&	\cellcolor{cgood}	&	\cellcolor{cbad}	\\	\hline \\
    \end{tabular}
    \tablefoot{Columns from left to right: model name, then best and worst case scenario for a source located in the direction of the Milky Way's bulk at varying distance, from $\unit[1]{kpc}$ to $\unit[200]{kpc}$. Green coloured cells means that at least 75\% of the post bounce points lie inside the 95\% confidence interval of the inferred ones, in yellow cells in which the recovered percentage is above 50\%, and in red cell where it is below 50\%.}
\end{table}

Due to their complex nature, \glspl{GW} from \glspl{CCSN} encode a significant amount of information. However, because of their weak intensity, a relatively nearby galactic event is necessary to capture most of this information. To infer the detectability of \glspl{GW}, we employ the pipeline \texttt{pnsInf} \citep{Bruel23}, which is designed to infer the properties of a \gls{PNS} from its waveform. We assess the detectability of two main quantities: (i) the time of bounce and (ii) the ratio $\sqrt{M/R^3}$ of the \gls{PNS}. 
Since the sky coverage of the detectors changes throughout the day due to Earth's rotation, we quantify how favourable the arrival time of an event is. We use the equivalent antenna pattern weighted by the nominal reach of each detector:
\begin{equation}
    \label{eq:antenna_pattern}
    F_\textnormal{eq} = \sqrt{ \frac{1}{\sum\limits_{k\in\textnormal{netw.}} w_k} \sum\limits_{k\in\textnormal{netw.}} w_k \left(F_+^k(\boldsymbol{\Omega}, t_0)^2 + F_\times^k(\boldsymbol{\Omega}, t_0)^2\right)},
\end{equation}
where $F_+^k$ and $F_\times^k$ are the detector response functions, $\boldsymbol{\Omega}$ is the position of the source in the sky (relative to an Earth-fixed coordinate system), $t_0$ is the GPS time of arrival (at the centre of the Earth), and $w_k$ represents the nominal reach of each detector.

Considering that most stars are located in the bulk of the Milky Way, we account for the sky coordinates and times of arrival for the best (worst) case scenario in the bulk direction (here taken as an ellipse centred at the galaxy centre of semi-minor and major axes of $\unit[15]{^\circ}$ and $\unit[20]{^\circ}$, respectively) where $F_\textnormal{eq}$ is maximum (minimum). From now on, we refer to the `best case' as the event with sky coordinates RA $\unit[16.99]{h}$ and Dec $\unit[-35.59]{^\circ}$ and time of arrival $\unit[1393213818]{s}$, and the `worst case' as the one with RA $\unit[17.48]{h}$, Dec $\unit[-13.99]{^\circ}$ at $\unit[1393238418]{s}$.

Prior to using the pipeline, we pad our waveforms with zeros before the bounce time and after the end of the simulation to avoid any boundary effects when calculating the time-frequency maps. Assuming a certain distance of the source from the detector network, we injected the waveforms into random Gaussian noise, which was different for every source and distance, and therefore modified the signal-to-noise of each event.

For the recovery of the time of bounce, we calculate the time-frequency maps for each detector in the network with a time window of $\unit[10]{ms}$. We then search for the brightest column of pixels in each detector in the frequency band $\unit[100-1100]{Hz}$, which are the typical frequencies present during the bounce. We consider the process successful if the recovered time lies within a $\unit[15]{ms}$ window of the actual time of bounce.

To infer $\sqrt{M/R^3}$ using the pipeline, we inverted the relation found in \cite{TorresForne19}.
We track the frequency of the likelihood calculated from the spectrograms computed with a $\unit[20]{ms}$ time window, starting from $\unit[25]{ms}$ after the bounce signal to avoid any signal arising from the bounce itself or strong prompt convection and \gls{PNS} core ringdown oscillations happening in the few milliseconds after the bounce. Finally, we compare it with the real ratio derived from the simulation.

For both analyses, we employ a source distance ranging from $\unit[1]{kpc}$ to $\unit[200]{kpc}$. While the bounce recovery is performed using the full interferometer network (LIGO, Virgo, and KAGRA), the mass-radius ratio inference uses just the two LIGO interferometers since they have the highest reach. The interferometer sensitivity is set to be the theoretical one for the O5 run \citep{LVK20}.

In Tables \ref{tab:bounce_det} and \ref{tab:mr_det}, we show the detection results obtained from the LHVK network for inferring the bounce time and the $\sqrt{M/R^3}$ ratio, respectively.
The bounce time can be well recovered until $\unit[30]{kpc}$ by at least one detector for almost every case and both scenarios. Additionally, in fast-rotating models or models in which resonance plays an important role, detection can be pushed up to $\unit[70]{kpc}$ for one detector, and for a subset of these latter ones to $\unit[100]{kpc}$. However, as shown in Section \ref{sec:bounce_convection}, slow rotators do not present a strong bounce signal. Nonetheless, we can recover a bounce time very close to the actual one. 

\section{General properties of our simulations}
\begin{figure*}
    \centering
    \includegraphics[width=17cm]{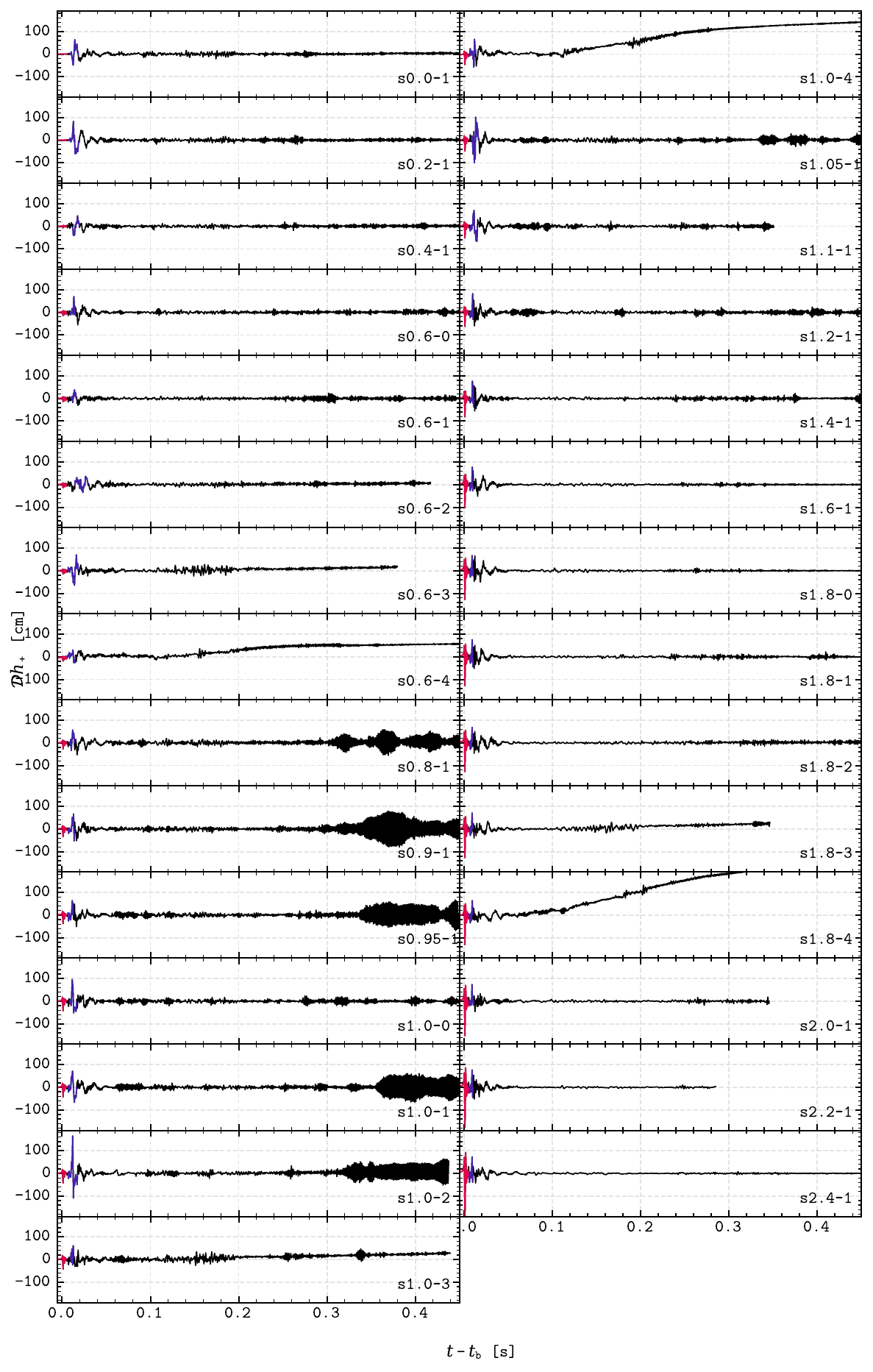}
    \caption{\gls{GW} amplitude evolution of the 29 models presented in this work. The red and blue portions  of the amplitude correspond to the bounce and prompt convection signal, respectively.}
    \label{fig:GWs_set}
\end{figure*}

In this section, we collect data related to our simulation sample.
Table \ref{tab:results} lists the quantitative results from our analysis and Figure \ref{fig:GWs_set} shows the waveforms of the models presented in this study.

\longtab[1]{
    \begin{landscape}
        \begin{table}
        \caption{\label{tab:results}Quantitative results.}
        \begin{tabular}{ccccccccccccccc}
            \multirow{2}{*}{Model}	& $T/|W|$ & $\Delta h_\textnormal{b}$ & $\Delta h_\textnormal{pb}$ & $\Delta h_\textnormal{core}$ & $\Delta h_\textnormal{conv,peak}$ & $f_\textnormal{b}$ & $f_\textnormal{pb}$ & $f_\textnormal{core}$ & $f_\textnormal{conv}$ & $E_\textnormal{GW,b}$ & $E_\textnormal{GW, pb}$ & $E_\textnormal{GW,300}$ & $E_\textnormal{GW}$ &  $t_\textnormal{expl}$ \\
             & $\left[10^{-2}\right]$ & [cm] & [cm] & [cm] & [cm] & [Hz] & [Hz] & [Hz] & [Hz] & [$10^{46}$ erg] & [$10^{46}$ erg] & [$10^{46}$ erg] & [$10^{46}$ erg]  & [s]\\\hline
            \texttt{s0.0-1} & 0.00 (0.00) & 5 & 114 & 50 & 64 & 1750 & 130 & 284 & 131 & 0.00 & 0.14 & 0.38 & 0.80&  0.15 \\
            \texttt{s0.2-1} & 0.03 (0.03) & 7 & 144 & 68 & 91 & 1220 & 120 & 876 & 117 & 0.00 & 0.25 & 0.50 & 1.31&  0.17 \\
            \texttt{s0.4-1} & 0.11 (0.11) & 19 & 88 & 36 & 52 & 650 & 130 & 634 & 118 & 0.01 & 0.12 & 0.35 & 0.84&  0.18 \\
            \texttt{s0.6-0} & 0.26 (0.28) & 32 & 88 & 63 & 51 & 600 & 100 & 435 & 98 & 0.03 & 0.28 & 0.36 & 1.41&  0.19 \\
            \texttt{s0.6-1} & 0.26 (0.29) & 33 & 70 & 30 & 39 & 610 & 120 & 452 & 111 & 0.03 & 0.11 & 0.31 & 1.04&  0.23 \\
            \texttt{s0.6-2} & 0.26 (0.27) & 32 & 72 & 41 & 47 & 600 & 90 & 514 & 99 & 0.03 & 0.15 & 0.38 & 0.75&  0.17 \\
            \texttt{s0.6-3} & 0.26 (0.28) & 31 & 136 & 51 & 66 & 600 & 150 & 662 & 134 & 0.03 & 0.27 & 0.39 & 0.76&  0.15 \\
            \texttt{s0.6-4} & 0.23 (0.23) & 31 & 56 & 36 & 35 & 600 & 220 & 668 & 106 & 0.02 & 0.11 & 0.61 & 1.06&  0.09 \\
            \texttt{s0.8-1}$^*$ & 0.41 (0.49) & 46 & 92 & 49 & 53 & 540 & 180 & 516 & 117 & 0.05 & 0.23 & 0.70 & 6.79&  0.17 \\
            \texttt{s0.9-1}$^*$ & 0.54 (0.61) & 52 & 122 & 95 & 59 & 500 & 620 & 661 & 122 & 0.07 & 0.41 & 0.86 & 19.23&  0.20 \\
            \texttt{s0.95-1}$^*$ & 0.63 (0.72) & 57 & 118 & 71 & 48 & 510 & 680 & 759 & 117 & 0.08 & 0.39 & 0.84 & 11.62&  0.20 \\
            \texttt{s1.0-0}$^*$ & 0.72 (0.76) & 61 & 149 & 100 & 70 & 510 & 160 & 331 & 135 & 0.09 & 0.36 & 0.68 & 1.13&  0.16 \\
            \texttt{s1.0-1}$^*$ & 0.72 (0.76) & 61 & 140 & 53 & 88 & 520 & 140 & 239 & 125 & 0.09 & 0.24 & 0.89 & 13.07&  0.18 \\
            \texttt{s1.0-2}$^*$ & 0.72 (0.79) & 61 & 276 & 216 & 91 & 520 & 700 & 625 & 144 & 0.09 & 1.15 & 0.60 & 9.44&  0.18 \\
            \texttt{s1.0-3} & 0.72 (0.75) & 61 & 95 & 76 & 34 & 520 & 720 & 740 & 136 & 0.09 & 0.47 & 0.77 & 1.33&  0.14 \\
            \texttt{s1.0-4} & 0.64 (0.60) & 63 & 125 & 103 & 47 & 550 & 370 & 839 & 180 & 0.09 & 0.31 & 0.64 & 1.10&  0.08 \\
            \texttt{s1.05-1}$^*$ & 0.73 (0.87) & 68 & 205 & 136 & 81 & 520 & 840 & 856 & 123 & 0.11 & 1.16 & 0.56 & 4.63&  0.16 \\
            \texttt{s1.1-1} & 0.77 (0.88) & 75 & 139 & 70 & 64 & 540 & 160 & 358 & 129 & 0.13 & 0.37 & 0.72 & 1.39&  0.19 \\
            \texttt{s1.2-1} & 0.96 (1.04) & 93 & 117 & 93 & 36 & 570 & 590 & 675 & 151 & 0.17 & 0.73 & 0.76 & 3.28&  0.19 \\
            \texttt{s1.4-1} & 1.24 (1.39) & 120 & 100 & 98 & 41 & 560 & 590 & 819 & 141 & 0.27 & 0.91 & 0.21 & 2.02&  0.21 \\
            \texttt{s1.6-1} & 1.59 (1.75) & 155 & 107 & 77 & 49 & 539 & 680 & 720 & 94 & 0.44 & 0.57 & 0.10 & 1.17&  0.20 \\
            \texttt{s1.8-0} & 2.02 (2.14) & 189 & 117 & 100 & 49 & 538 & 630 & 616 & 86 & 0.68 & 0.69 & 0.10 & 1.50&  0.21 \\
            \texttt{s1.8-1} & 2.03 (2.11) & 189 & 121 & 105 & 46 & 543 & 650 & 652 & 97 & 0.68 & 0.61 & 0.12 & 1.59&  0.17 \\
            \texttt{s1.8-2} & 2.02 (2.13) & 189 & 116 & 86 & 44 & 538 & 650 & 632 & 100 & 0.68 & 0.55 & 0.12 & 1.81&  0.18 \\
            \texttt{s1.8-3} & 2.02 (2.14) & 188 & 112 & 67 & 39 & 542 & 680 & 745 & 110 & 0.68 & 0.35 & 0.16 & 1.25&  0.14 \\
            \texttt{s1.8-4} & 2.19 (1.83) & 186 & 100 & 75 & 27 & 540 & 720 & 636 & 75 & 0.65 & 0.26 & 0.74 & 2.05&  0.07 \\
            \texttt{s2.0-1} & 2.63 (2.57) & 229 & 122 & 75 & 30 & 540 & 480 & 568 & 171 & 1.03 & 0.39 & 0.12 & 1.59&  0.19 \\
            \texttt{s2.2-1} & 2.85 (2.97) & 268 & 124 & 109 & 30 & 536 & 580 & 587 & 75 & 1.36 & 0.61 & 0.05 & 2.02&  0.19 \\
            \texttt{s2.4-1} & 3.65 (3.39) & 295 & 114 & 95 & 39 & 525 & 550 & 555 & 90 & 1.69 & 0.35 & 0.04 & 2.10&  0.20 \\ \hline
            \end{tabular}
          \tablefoot{Columns from left to right provide: name of the simulation; ratio between \gls{PNS} core and convective envelope kinetic rotational energy and gravitational energy at bounce ($T/|W|$), and at $\unit[15]{ms}$ in brackets; difference of highest and lowest point of the waveform for bounce ($\Delta h_\textnormal{b}$), post-bounce ($\Delta h_\textnormal{pb}$), core ($\Delta h_\textnormal{core}$), and convection signals ($\Delta h_\textnormal{conv,peak}$); peak frequency for bounce ($f_\textnormal{b}$), post bounce ($f_\textnormal{pb}$), core($f_\textnormal{core}$), and convection signals ($f_\textnormal{conv}$); energy carried away as \glspl{GW} for the  bounce signal $E_\textnormal{GW,b}$, post-bounce signal $E_\textnormal{GW,pb}$, from $\unit[25]{ms}$ until $\unit[300]{ms}$, $E_\textnormal{GW,300}$, and total \glspl{GW} energy $E_\textnormal{GW}$; explosion time, $t_\textnormal{expl}$. Models in which the \gls{GW} signal presents resonance are marked with $^*$. }
        \end{table}
    \end{landscape}
}

\end{appendix}
\label{LastPage}
\end{document}